\documentclass[preprint, aps,prd,amsmath,amssymb,floatfix,nofootinbib]{revtex4}
\usepackage{graphicx}
\usepackage{color}
\usepackage{graphicx}
\usepackage{dcolumn}
\usepackage{bm}
\usepackage{amsmath}
\usepackage{amsfonts}
\usepackage{bbm}
\usepackage{subfigure}
\usepackage{pxfonts}

\usepackage{setspace}
\usepackage{slashed}
\newcommand{\beq}{\begin{eqnarray}}
\newcommand{\eeq}{\end{eqnarray}}

\newcommand{\bmp}{\noindent\begin{minipage}{16cm}}
\newcommand{\emp}{\end{minipage}\vskip 7mm} 

\def\drawbox#1#2{\hrule height#2pt
        \hbox{\vrule width#2pt height#1pt \kern#1pt
              \vrule width#2pt}
              \hrule height#2pt}

\def\Asym#1#2{\vcenter{\vbox{\drawbox{#1}{#2}
              \kern-#2pt 
              \drawbox{#1}{#2}}}}

\def\mpt{{\slash\!\!\!\!\!\:p}_T}
\def\mptv{{\slash\!\!\!\!\!\:\vec{p}}_T}

\definecolor{rossoCP3}{cmyk}{0,.88,.77,.40}

\begin{document}

\title{\Large  \color{rossoCP3}    Vanilla Technicolor at Linear Colliders }
\author{Mads T. {\sc Frandsen}$^{\color{rossoCP3}{\clubsuit}}$}\email{m.frandsen1@physics.ox.ac.uk} 
 \author{Matti  {\sc J\"arvinen}$^{\color{rossoCP3}{\spadesuit}}$}\email{mjarvine@physics.uoc.gr} 
\author{Francesco Sannino$^{\color{rossoCP3}{\varheartsuit}}$}\email{sannino@cp3-origins.net} 
\affiliation{\mbox{$^{\color{rossoCP3} {\clubsuit}}$Rudolf Peierls
  Centre for Theoretical Physics, Univ. of Oxford, UK} \\
 \mbox{$^{\color{rossoCP3} {\spadesuit}}$ Crete Center for Theoretical Physics, Dept. of Physics, Univ. of Crete, 71003 Heraklion, Greece} \\
\mbox{$^{\color{rossoCP3} {\varheartsuit}}${ CP}$^{ \bf 3}${-Origins}, 
 Univ. of Southern Denmark, Campusvej 55, DK-5230 Odense M, Denmark}}
 
\begin{flushright}
{\it CCTP-2011-09}\\
{\it CP$^\mathit{3}$-Origins-2011-08}
\end{flushright}

\begin{abstract} 
We analyze the reach of Linear Colliders (LC)s for models of dynamical electroweak symmetry breaking. We show that LCs  can efficiently test the compositeness scale, identified with the mass of the new spin-one resonances, till the maximum energy in the center-of-mass of the colliding leptons.  In particular we analyze the Drell-Yan processes involving spin-one intermediate heavy bosons decaying either leptonically or into two Standard Model (SM) gauge bosons. We also analyze the light Higgs production in association with a SM gauge boson stemming also from an intermediate spin-one heavy vector.\end{abstract}

\maketitle

\section{Vanilla Technicolor setup for Linear Colliders} \label{sec:EffL}

The Large Hadron Collider (LHC) is producing a wealth of experimental results which are already providing interesting constraints for time-honored extensions of the SM of high energy physics. It is therefore timely to explore, in case a dynamical origin of the electroweak symmetry breaking occurs, the benefits stemming from the construction of future LCs.

Based on recent progress 
\cite{Sannino:2004qp,Dietrich:2006cm,Ryttov:2007sr,Ryttov:2007cx,Dietrich:2005jn,Sannino:2008ha} in the
understanding of Walking Technicolor (WT)  dynamics \cite{Holdom:1981rm,Holdom:1984sk,Eichten:1979ah,Lane:1989ej} various phenomenologically viable models have been proposed. Primary examples are: i) the $SU(2)$ theory with two techniflavors in the adjoint representation,  known as Minimal Walking Technicolor (MWT) \cite{Sannino:2004qp}; ii) the $SU(3)$ theory with
two flavors in the two-index symmetric representation which is called Next to Minimal
Walking Technicolor (NMWT) \cite{Sannino:2004qp} and iii) The $SU(2)$ theory with two techniflavors in the fundamental representation and 1 techniflavor in the adjoint representation known as (UMT) \cite{Ryttov:2008xe}. These gauge theories have remarkable 
properties~\cite{Sannino:2004qp,Hong:2004td,Dietrich:2005jn,Dietrich:2006cm,Ryttov:2007sr,Ryttov:2007cx} and alleviate the tension with the LEP precision data when used for Technicolor~\cite{Sannino:2004qp,Dietrich:2005jn,Foadi:2007ue,Foadi:2007se,Foadi:2008xj}. The effects of the extensions of the Technicolor models to be able to account for the SM fermion masses are important and cannot be neglected as shown in \cite{Fukano:2010yv}.

Despite the different envisioned underlying gauge dynamics it is a fact that the SM structure alone requires the extensions to contain, at least, the following chiral symmetry breaking pattern (insisting on keeping the custodial symmetry of the SM): 
\beq
\label{basepattern}
SU(2)_{\rm L}\times SU(2)_{\rm R}  \to SU(2)_{\rm V} \ .
\eeq

We will call this common sector of any Technicolor extension of the SM, the {\it vanilla} sector. The reason for such a name is that the vanilla sector is common to old models of Technicolor featuring running dynamics and the ones featuring walking dynamics associated to a slow running of the Technicolor gauge coupling constant. It is worth mentioning that the {\it vanilla} sector is common not only to Technicolor extensions but to {\it any} known extension, even of extra-dimensional type, in which the Higgs sector can be viewed as composite. In fact, the effective Lagrangian we are about to introduce can  be used for modeling several extensions with a common {vanilla} sector respecting the same constraints spelled out in \cite{Foadi:2007ue}. The natural candidate for a walking technicolor model featuring exactly this global symmetry is NMWT \cite{Sannino:2004qp}.

Based on the {\it vanilla symmetry} breaking pattern we describe the low energy spectrum in terms of the lightest spin-one vector and axial-vector iso-triplets $V^{\pm,0}, A^{\pm,0}$ as well as the lightest iso-singlet scalar resonance $H$. In QCD the equivalent states are the $\rho^{\pm,0}$, $a_1^{\pm}$ and the $f_0(600)$ \cite{Belyaev:2008yj}. The 3 technipions $\Pi^{\pm ,0}$ produced in the symmetry breaking become the longitudinal components of the $W$ and $Z$ bosons.

The composite spin-one and spin-0 states and their interaction with the SM fields are described via the following effective Lagrangian which we developed, first for minimal models of walking technicolor  \cite{Foadi:2007ue,Appelquist:1999dq}: 
\begin{eqnarray}
{\cal L}_{\rm boson}&=&-\frac{1}{2}{\rm Tr}\left[\widetilde{W}_{\mu\nu}\widetilde{W}^{\mu\nu}\right]
-\frac{1}{4}\widetilde{B}_{\mu\nu}\widetilde{B}^{\mu\nu}
-\frac{1}{2}{\rm Tr}\left[F_{{\rm L}\mu\nu} F_{\rm L}^{\mu\nu}+F_{{\rm R}\mu\nu} F_{\rm R}^{\mu\nu}\right] \nonumber \\
&+& m^2\ {\rm Tr}\left[C_{{\rm L}\mu}^2+C_{{\rm R}\mu}^2\right]
+\frac{1}{2}{\rm Tr}\left[D_\mu M D^\mu M^\dagger\right]
-\tilde{g^2}\ r_2\ {\rm Tr}\left[C_{{\rm L}\mu} M C_{\rm R}^\mu M^\dagger\right] \nonumber \\
&-&\frac{i\ \tilde{g}\ r_3}{4}{\rm Tr}\left[C_{{\rm L}\mu}\left(M D^\mu M^\dagger-D^\mu M M^\dagger\right)
+ C_{{\rm R}\mu}\left(M^\dagger D^\mu M-D^\mu M^\dagger M\right) \right] \nonumber \\
&+&\frac{\tilde{g}^2 s}{4} {\rm Tr}\left[C_{{\rm L}\mu}^2+C_{{\rm R}\mu}^2\right] {\rm Tr}\left[M M^\dagger\right]
+\frac{\mu^2}{2} {\rm Tr}\left[M M^\dagger\right]-\frac{\lambda}{4}{\rm Tr}\left[M M^\dagger\right]^2
\label{eq:boson}
\end{eqnarray}
where $\widetilde{W}_{\mu\nu}$ and $\widetilde{B}_{\mu\nu}$ are the ordinary electroweak field strength tensors, $F_{{\rm L/R}\mu\nu}$ are the field strength tensors associated to the vector meson fields $A_{\rm L/R\mu}$~\footnote{In Ref.~\cite{Foadi:2007ue}, where the chiral symmetry is SU(4), there is an additional term whose coefficient is labeled $r_1$. With an SU($N$)$\times$SU($N$) chiral symmetry this term is just identical to the $s$ term.}, and the $C_{{\rm L}\mu}$ and $C_{{\rm R}\mu}$ fields are
\begin{eqnarray}
C_{{\rm L}\mu}\equiv A_{{\rm L}\mu}^a T^a-\frac{g}{\tilde{g}}\widetilde{W_\mu^a} T^a\ , \quad
C_{{\rm R}\mu}\equiv A_{{\rm R}\mu}^a T^a-\frac{g^\prime}{\tilde{g}}\widetilde{B_\mu} T^3\ ,
\end{eqnarray}
where $T^a=\sigma^a/2$, and $\sigma^a$ are the Pauli matrices.The 2$\times$2 matrix $M$ is
\begin{eqnarray}
M=\frac{1}{\sqrt{2}}\left[v+H+2\ i\ \pi^a\ T^a\right]\ ,\quad\quad  a=1,2,3
\end{eqnarray}
where $\pi^a$ are the Goldstone bosons produced in the chiral symmetry breaking, $v=\mu/\sqrt{\lambda}$ is the corresponding VEV, and $H$ is the composite Higgs. The covariant derivative is
\begin{eqnarray}
D_\mu M&=&\partial_\mu M -i\ g\ \widetilde{W}_\mu^a\ T^a M + i\ g^\prime \ M\ \widetilde{B}_\mu\ T^3\ . 
\end{eqnarray}
When $M$ acquires its VEV, the Lagrangian of Eq.~(\ref{eq:boson}) contains mixing matrices for the spin-one fields. The mass eigenstates are the ordinary SM bosons, and two triplets of heavy mesons, of which the lighter (heavier) ones are denoted by $R_1^\pm$ ($R_2^\pm$) and $R_1^0$ ($R_2^0$). These heavy mesons are the only new particles, at low energy, relative to the SM.

Now we must couple the SM fermions. 
The minimal form for the quark Lagrangian is 
\begin{eqnarray}
{\cal L}_{\rm quark}&=&\bar{q}^i_L\ i \slashed{D} q_{iL} + \bar{q}^i_R\ i \slashed{D} q_{iR} \nonumber \\
&-&\left[\bar{q}^i_L\ (Y_u)_i^j\ M\ \frac{1+\tau^3}{2}\ q_{jR}
+\bar{q}^i_L\ (Y_d)_i^j\ M \ \frac{1-\tau^3}{2}\ q_{jR} + {\rm h.c.}\right] \ ,
\label{eq:quark}
\end{eqnarray}
where $i$ and $j$ are generation indices, and $i=1,2,3$, $q_{iL/R}$ are electroweak doublets.
The covariant derivatives are the ordinary
electroweak ones,
\begin{eqnarray}
\slashed{D}q_{iL}&=&\left(\slashed{\partial}-i\ g\ \slashed{\widetilde{W}}^a\ T^a
-i\ g^\prime \slashed{\widetilde{B}} Y_{\rm L}\right)q_{iL} \ ,\nonumber \\
\slashed{D}q_{iR}&=&\left(\slashed{\partial}-i\ g^\prime \slashed{\widetilde{B}} Y_{\rm R}\right)q_{iR} \ ,
\end{eqnarray}
where $Y_{\rm L}=1/6$ and $Y_{\rm R}={\rm diag}(2/3,-1/3)$. As usual, one can exploit the global symmetries of the kinetic terms 
to reduce the number of physical parameters in the Yukawa matrices $Y_u$ and $Y_d$. Thus we can take
\begin{eqnarray}
Y_u={\rm diag}(y_u,y_c,y_t) \ , \quad Y_d= V\ {\rm diag}(y_d,y_s,y_b) \ ,
\end{eqnarray}
and
\begin{eqnarray}
q^i_L=\left(\begin{array}{c} u_{iL} \\ V_i^j d_{jL} \end{array}\right) \ , \quad
q^i_R=\left(\begin{array}{c} u_{iR} \\ d_{iR} \end{array}\right) \ ,
\end{eqnarray}
where $V$ is the CKM matrix. 
One can also add mixing terms of the fermions with the $C_{\rm L}$ and $C_{\rm R}$ fields \cite{Foadi:2007ue}. We will however neglect them in our analysis, since they affect the tree-level anomalous couplings highly constrained by experiments.

\subsection{Parameter Space of Vanilla Technicolor}

Some of the parameters of the tree-level Lagrangian can be related to the electroweak $S$-parameter and to the masses of the composite particles as shown in \cite{Foadi:2007ue,Belyaev:2008yj}.
{}To be concrete we assume here $S\simeq 0.3$ corresponding approximately to its naive value in the 
NMWT model. The remaining parameters are the tree-level mass of the axial spin-one state $M_A$, the technicolor interaction strength $\tilde{g}$, the coupling $s$, and the composite Higgs mass $M_H$. The two parameters $s$ and $M_H$  mostly impact processes involving the composite Higgs. 
\begin{figure}
\includegraphics[height=8cm,width=10cm]{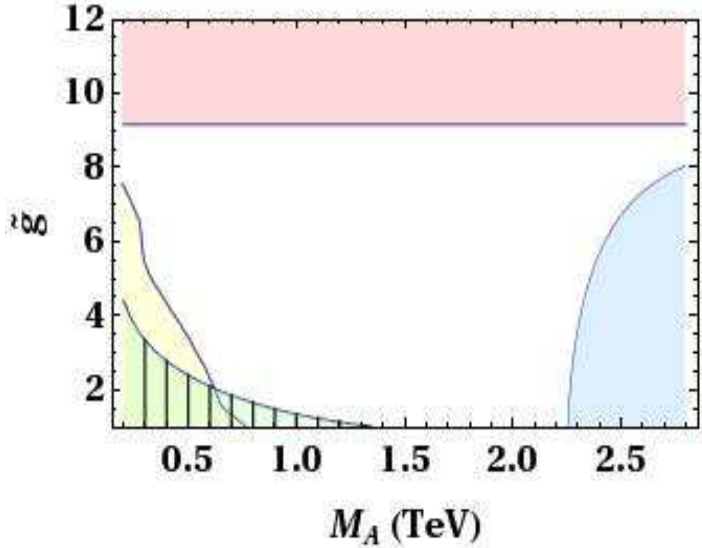}
\caption{Bounds, for $S=0.3$, in the $(M_A,\tilde{g})$ plane from: (i) CDF
direct searches of $R_1^0$ at Tevatron, in $p\bar{p}\rightarrow e^+e^-$, for
$s=0$ and  $M_{H}=200$~GeV.
The forbidden regions is the uniformly shaded one in the left corner.
(ii) Measurement of the electroweak parameters W and Y at
95\% confidence level. The forbidden region is the striped one in
the left corner. (iii) The excluded region in the right-lower corner comes from imposing the modified Weinberg's sum rules.  (iv) Consistency of the theory, i.e. reality for the vector and axial decay constants leads to  the horizontal stripe in the upper  part
of the figure.  We stress again that  
the shaded regions are excluded.}
\label{fig:bounds}
\end{figure}
Bounds on these parameters come both, from the electroweak precision tests, and direct searches. We shall give here a brief review of the constraints discussed in detail in \cite{Belyaev:2008yj} and not yet improved by the recent LHC experiments.
CDF imposes~\cite{Aaltonen:2008vx} lower bounds on $M_A$ and $\tilde{g}$ from direct searches of
$R_1^0$ in  the $p\bar{p}\rightarrow e^+e^-$ process, as shown by {the uniformly shaded region} 
in the lower left of Fig~\ref{fig:bounds}. 
The measurements of the electroweak parameters $W$ and $Y$ exclude~\cite{Foadi:2007se} the striped region 
on the lower left corner shown in Fig~\ref{fig:bounds} at the 95\% confidence level. The upper bound for $\tilde{g}$, shown by the upper horizontal line in Fig~\ref{fig:bounds}, is dictated by the internal consistency of the model.  The upper bound for $M_A$ corresponds to the value for
which both Weinberg's sum rules are satisfied in the running and walking regime of Technicolor \cite{Appelquist:1998xf,Foadi:2007ue}. This bound is shown in the lower right corner of Fig~\ref{fig:bounds}.

Recently, additional bounds from unitarity have been studied in \cite{Foadi:2008xj}, and the interesting possibility of using flavor data to constrain directly the technicolor sector has also been considered in \cite{Fukano:2009zm}. We have checked that these constraint do not further reduce the parameter space in our case.

\subsection{Key features of Vanilla Technicolor Phenomenology at LCs}
When turning off the electroweak interactions interesting decay modes of the spin-one massive vector states and composite scalar are: \beq
V \to \Pi\ \Pi \ , \quad A \to H\ \Pi \ , \quad H \to \Pi\ \Pi \ ,
\label{basedecays}
\eeq
with the appropriate charge assignments. We assumed here that the composite Higgs is lighter than the vector states. This is the case for the scalar field in QCD \cite{Sannino:1995ik,Harada:1995dc,Harada:1996wr,Harada:2003em}.
Once the electroweak interactions are turned on the technipions become the longitudinal components of the $W$ and $Z$ bosons. Therefore the processes in \eqref{basedecays} allow to detect the spin-one resonances at LCs. Here we are making the assumption that $2M_Z < M_H < M_{A,V}$. For the neutral vectors:
\beq
V \to W W \ , \quad vs \quad A \to H\ Z \ .
\eeq
This picture is not quite complete since the massive spin-one states mix with the SM gauge bosons. After diagonalizing the spin-one mixing matrices (see \cite{Foadi:2007ue,Belyaev:2008yj}) the lightest and heaviest of the composite spin-one triplets are termed $R_{1}^{\pm, 0}$ and $R_{2}^{\pm, 0}$ respectively. In the region of parameter space where $R_{1}$ is mostly an axial-like vector (for a mass less than or about one TeV) and $R_2$ mostly a vector state has the following qualitative dependence of the couplings to the SM fields as function of the electroweak gauge coupling $g$ and $\tilde{g}$:
\beq \label{eq:couplbeh}
g_{R_{1,2} f\bar{f}}\sim \frac{g}{\tilde{g}}  \ , \quad g_{R_2 W W}\sim \tilde{g} \ , \quad g_{R_1 H Z}\sim \tilde{g} \ .
\eeq
Notice that, since the heavy spin-one states do not couple directly to SM fermions, the couplings $g_{R_{1,2} f\bar{f}}$ arises solely from the mixing with $W$ and $Z$. This coupling is roughly proportional to $g/\tilde{g}$.

Using as guidance the picture above together with the mass spectrum inequality $M_H < M_{R_1} < M_{R_2}$ we set up the following collider search strategy for the heavy vectors and the composite Higgs.  
For small $\tilde{g}$ (meaning $g \lesssim \tilde{g} \lesssim 2$)  the coupling of $R_{1,2}$ to fermions is large and therefore the heavy spin-one states are produced directly via the elementary process $e^+e^- \to R_{1,2}$. 
In this regime the couplings $R_{1,2}$ to fermions are roughly equal and therefore it is easy to identify these states as peaks  in the di-lepton final state distributions. In the large $\tilde g$ regime ($ \tilde{g} > 2$) from \eqref{eq:couplbeh} it is clear that it is better to consider the direct production of the heavy states followed by decays to SM gauge bosons as well as one SM gauge boson and a composite Higgs, i.e.  
$e^+e^- \to R_1 \to HZ $ and $e^+e^- \to R_2 \to W^+W^-$.

\section{Phenomenology}\label{sec:pheno}

To perform the signal and background analysis 
we use the CalcHEP~\cite{Pukhov:2004ca} implementation\footnote{The FeynRules implementation can be downloaded here https://feynrules.phys.ucl.ac.be/wiki/TechniColor and here http://cp3-origins.dk/research/tc-tools.} of the above model described in \cite{Belyaev:2008yj}. The LanHEP
package~\cite{Semenov:2008jy} has been used to derive the Feynman rules for
the model.  CalcHEP \cite{calchep_man} implements the Jadach, Skrzypek and Ward expressions of Ref.~\cite{ISR} for Initial State Radiation (ISR) and we use the 
the parameterisation of Beamstrahlung specified for the 
International LC
(ILC) project in \cite{ILC_RDR}:
\begin{eqnarray}
\mbox{Horizontal beam  size (nm)} &=& 640,   \nonumber \\
\mbox{Vertical   beam  size (nm)} &=& 5.7,   \nonumber \\
\mbox{Bunch length (mm)}          &=& 0.300, \nonumber \\
\mbox{Number of particles in the bunch (N)} &=& 2\times10^{10}.
\end{eqnarray}
These parameters were recently employed in a study of dimuon \cite{Basso:2009hf} and Higgs \cite{Basso:2010si} production at LCs in $Z'$ models.

We shall consider two center-of-mass energies, $\sqrt{s} = 1$ and 3~TeV, corresponding to the maximal energies of ILC and CLIC, respectively. 
Within the allowed parameter space of vanilla technicolor it is possible to identify two limiting regions \cite{Foadi:2007ue,Foadi:2008xj} which we term here the {\it low-mass} and {\it high-mass} regions of the spin-one vector mesons. These are the regions  investigated here. 

In the low-mass region, the axial resonance is lighter than the vector one, and both resonance masses are below one TeV. On the other hand, in the heavy mass region the vector resonance is lighter than the axial, and the masses are above $1.5$ to $2$~TeV.
A LC with $\sqrt{s}=1$~TeV, as we shall see, provides a detailed study of vanilla technicolor in the low-mass region, whereas the high-mass region can be accessed with $\sqrt{s}=3$~TeV. 

To be concrete we choose as reference values of the tree-level mass scale $M_A$ to be around $750$~GeV and $2250$~GeV corresponding to the low mass and high mass cases respectively. The physical masses of the heavy spin-one states are reported in Table \ref{table:masses}. 
The mass difference of the physical eigenstates $R_1$ (for low masses) and $R_2$ (for high masses) with respect to $M_A$ is due mostly to the mixing with the SM gauge bosons.
\begin{table}
\begin{eqnarray}
\begin{array}{c|c||cc|cc}
 M_A(\textrm{GeV})  & \quad \ \tilde g \quad \mbox{} &\quad M_{R_1^0}(\textrm{GeV})  &\quad \Gamma_{R_1^0}(\textrm{GeV})  &\quad M_{R_2^0}(\textrm{GeV})  &\quad \Gamma_{R_2^0}(\textrm{GeV}) \\
\hline\hline
750 &  \quad \ 2  \quad \mbox{} & \quad  773 &\quad 1.15  &\quad 844 &\quad 2.53 \\
 &  \quad \ 3 \quad \mbox{} & \quad  760 &\quad 0.813  &\quad 894 &\quad 6.32 \\
 &  \quad \ 5 \quad \mbox{} & \quad  753 &\quad 1.81  &\quad 1080 &\quad 82.0 \\
 &  \quad \ 8 \quad \mbox{} & \quad  750 &\quad 16.2  &\quad 1440 &\quad 265 \\
\hline
2250 &  \quad \ 2  \quad \mbox{} & \quad  2270 &\quad 29.6  &\quad 2360 &\quad 36.1\\
 & \quad \ 3 \quad \mbox{} & \quad  2220 &\quad 60.4  &\quad 2290 &\quad 64.0 \\ 
 & \quad \ 5 \quad \mbox{} & \quad  2090 &\quad 127  &\quad 2260 &\quad 170 \\ 
 & \quad \ 8 \quad \mbox{} & \quad  1770 &\quad 148  &\quad 2250 &\quad 425 
\end{array} \nonumber
\end{eqnarray}
\caption{Physical masses and decay widths of the heavy vector mesons.}
\label{table:masses}
\end{table}

We also assume the composite Higgs mass to be around $M_H=200$~GeV unless stated otherwise. Note that the Higgs mass affects the $e^+e^- \to \ell^+\ell^-$ and $e^+e^- \to W^+W^-$ processes via the widths of the heavy technivector states.

\begin{figure}[tbh]
\includegraphics[width=0.45\textwidth]{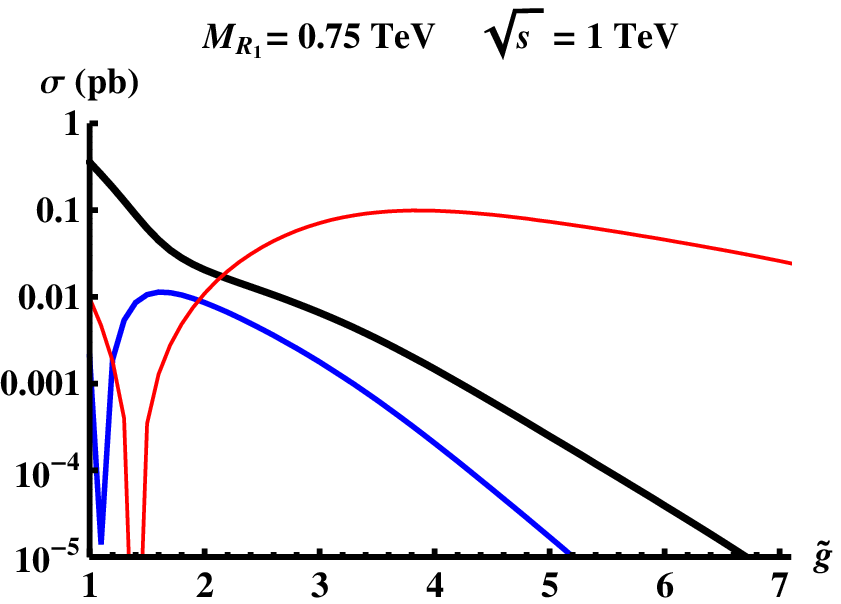}%
\hskip .5cm
\includegraphics[width=0.45\textwidth]{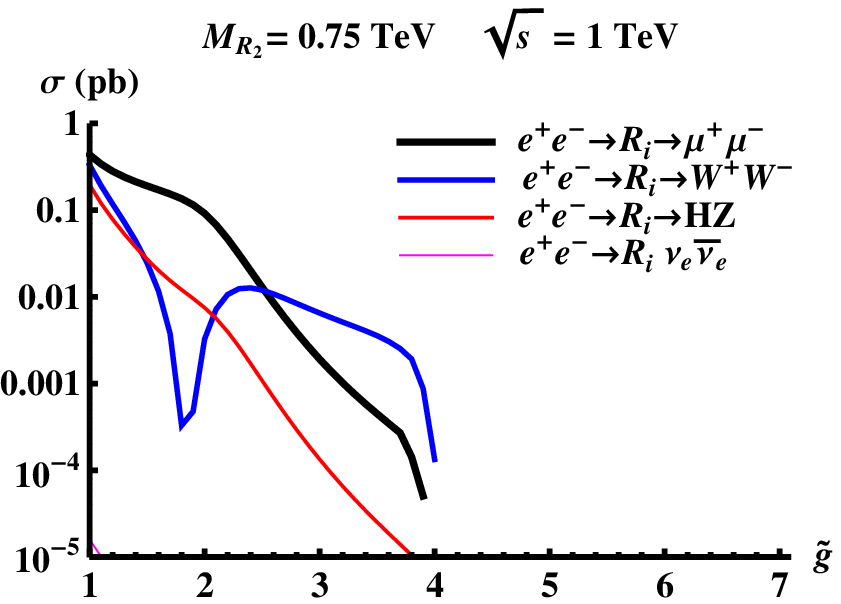} 

\vspace{10mm}

\includegraphics[width=0.45\textwidth]{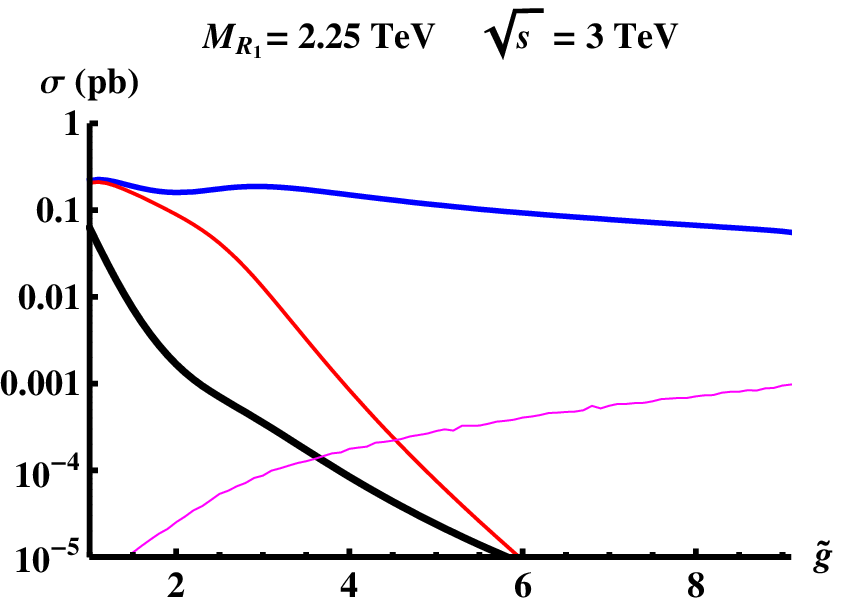}%
\hskip .5cm
\includegraphics[width=0.45\textwidth]{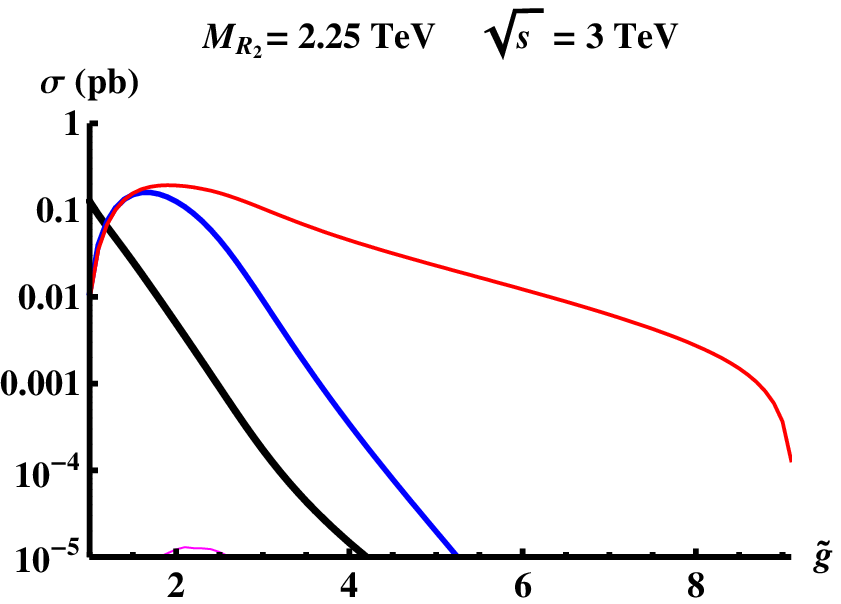}%
\caption{\label{fig:gtdep}Top row: The signal cross-sections for key processes  at $\sqrt{s} = 1$~TeV as a function of $\tilde g$ with keeping the resonance masses $M_{R_1}$ (left) and $M_{R_2}$ (right) fixed at $0.75$~TeV. Bottom row: The same for $\sqrt{s} = 3$~TeV, with masses fixed at $2.25$~TeV.  }
\end{figure}

We start by plotting in Fig.~\ref{fig:gtdep} the cross sections associated to the signals for the selected processes as a function of $\tilde g$. For the dimuon, $W^+W^-$ and $HZ$ final states we included only the diagrams with the heavy spin-one resonance $R_1^0$ ($R_2^0$) as intermediate states. The plots on the left (right) correspond to $R_1^0$ ($R_2^0$). We use two reference values for the center-of-mass energy, $\sqrt{s} = 1$ (top row) and $3$~TeV (bottom row). Notice that we keep the physically relevant resonance  mass ($M_{R_1}$ on the left and $M_{R_2}$ on the right)  fixed. 

Albeit the detailed cross section differ for different processes it is clear from Fig.~\ref{fig:gtdep} that the dimuon and diboson cross sections are of the same order of magnitude for sufficiently low $\tilde g$. This is the region where the axial vector and vector spin-one states are heavily mixed. The dimuon process $e^+e^- \to R_{1,2}^0 \to \mu^+\mu^-$ (black thick lines) has the largest cross section for small $\tilde g$ and for $\sqrt{s} = 1$~TeV. For $\sqrt{s}=3$~TeV the cross sections for the diboson channels $e^+e^- \to R_{1,2}^0 \to W^+W^-$ (blue lines) and $e^+e^- \to R_{1,2}^0 \to HZ$ (red lines) are enhanced, but the dimuon channel is still expected to produce the best signal for low $\tilde g$ since it has a cleaner final state. For large $\tilde g$ the $HZ$ final state (for axial vector spin-one resonances) and the $W^+W^-$ final state (for vector spin-one resonances) are dominant as expected. 

Fig.~\ref{fig:gtdep} also includes the cross section for the vector boson fusion  $e^+e^- \to R_{1,2}^0 +2 \nu$ (thin magenta lines). In principle this process would be interesting at large $\tilde{g}$ because the heavy vector boson is produced via the fusion of two $W$s and therefore the production vertex depends only on $\tilde{g}$. However, due to the dynamical constraints for running and walking technicolor imposed in \cite{Foadi:2007ue}  this process turns out to be suppressed with respect to the diboson  processes. We will, therefore, analyze in detail the following signatures:
\begin{itemize}
\item[(1)]
 $e^+ e^-\to R^{0}_{1,2}\to \ell^+\ell^-$
\item[(2)]
  $e^+ e^-\to R^{0}_{1,2}\to W^\pm W^\mp\to 
 \ell + \nu + 2j$
\item[(3)]
  $e^+ e^-\to R^{0}_{1,2}\to ZH\to 
\ 2\ell 4 j$
,
\end{itemize}
where $\ell$ denotes a charged lepton (electron or muon)  and $j$ denotes a jet. We choose as reference values $\tilde g=2,\ 3, \ 5$, and $8$. As pointed out above, and as we shall see in the detailed analysis below, signature (1) is expected to yield the best signal for $\tilde g=2, 3$, while (2) and (3) are relevant for investigating the values $\tilde g=5,8$.

\begin{figure}[htb!]
\includegraphics[width=0.45\textwidth]{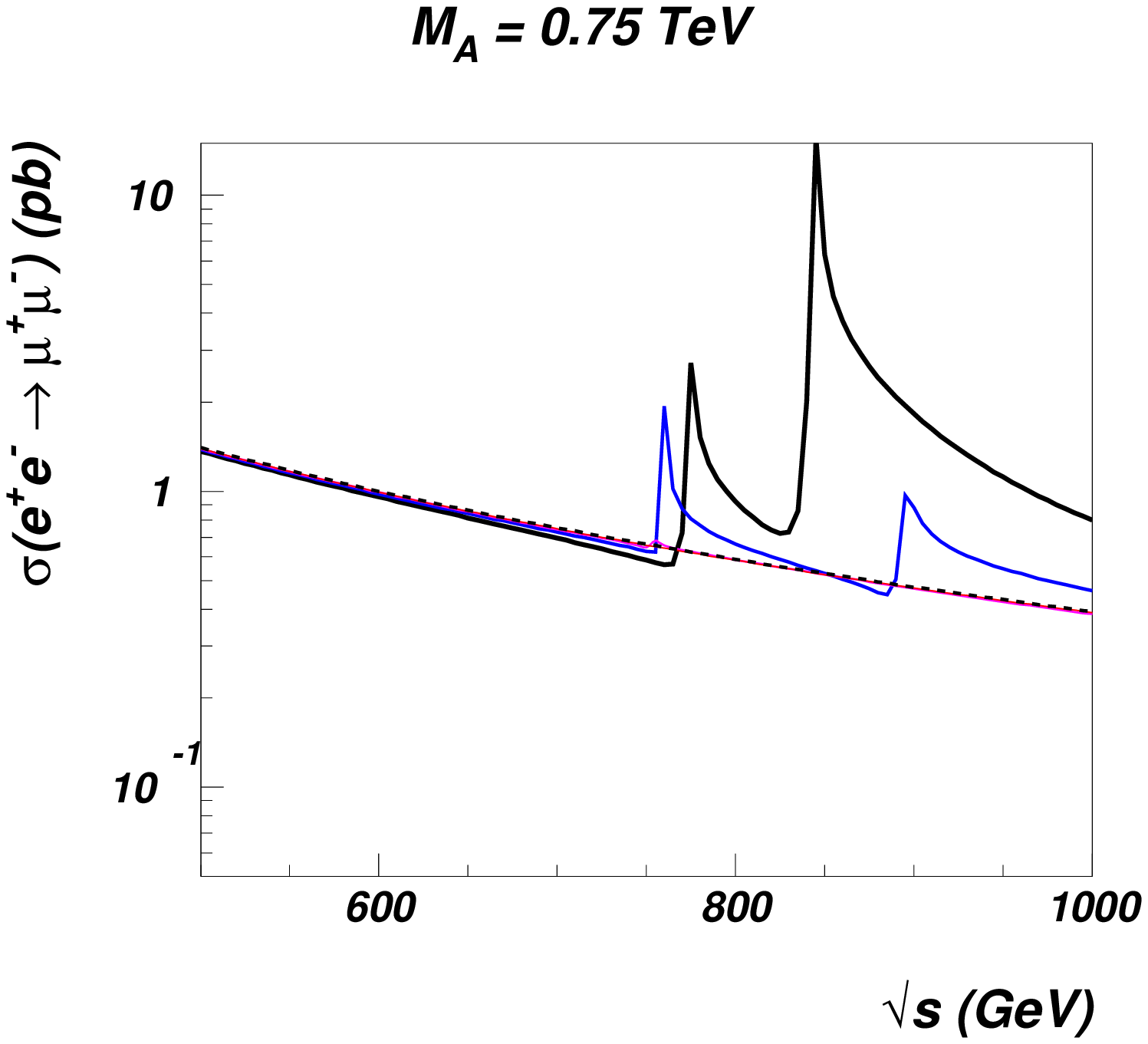}%
\includegraphics[width=0.45\textwidth]{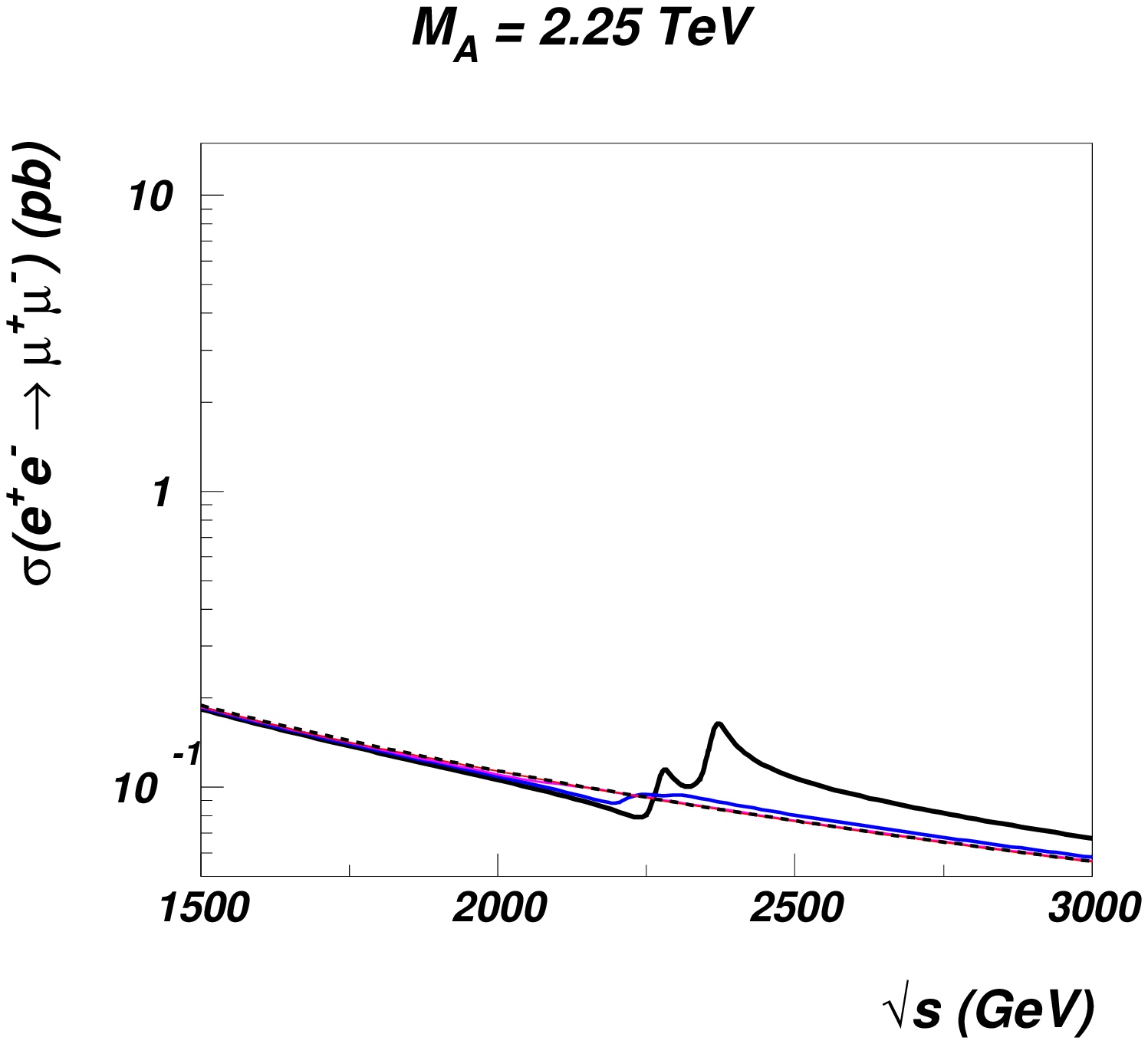}
\includegraphics[width=0.45\textwidth]{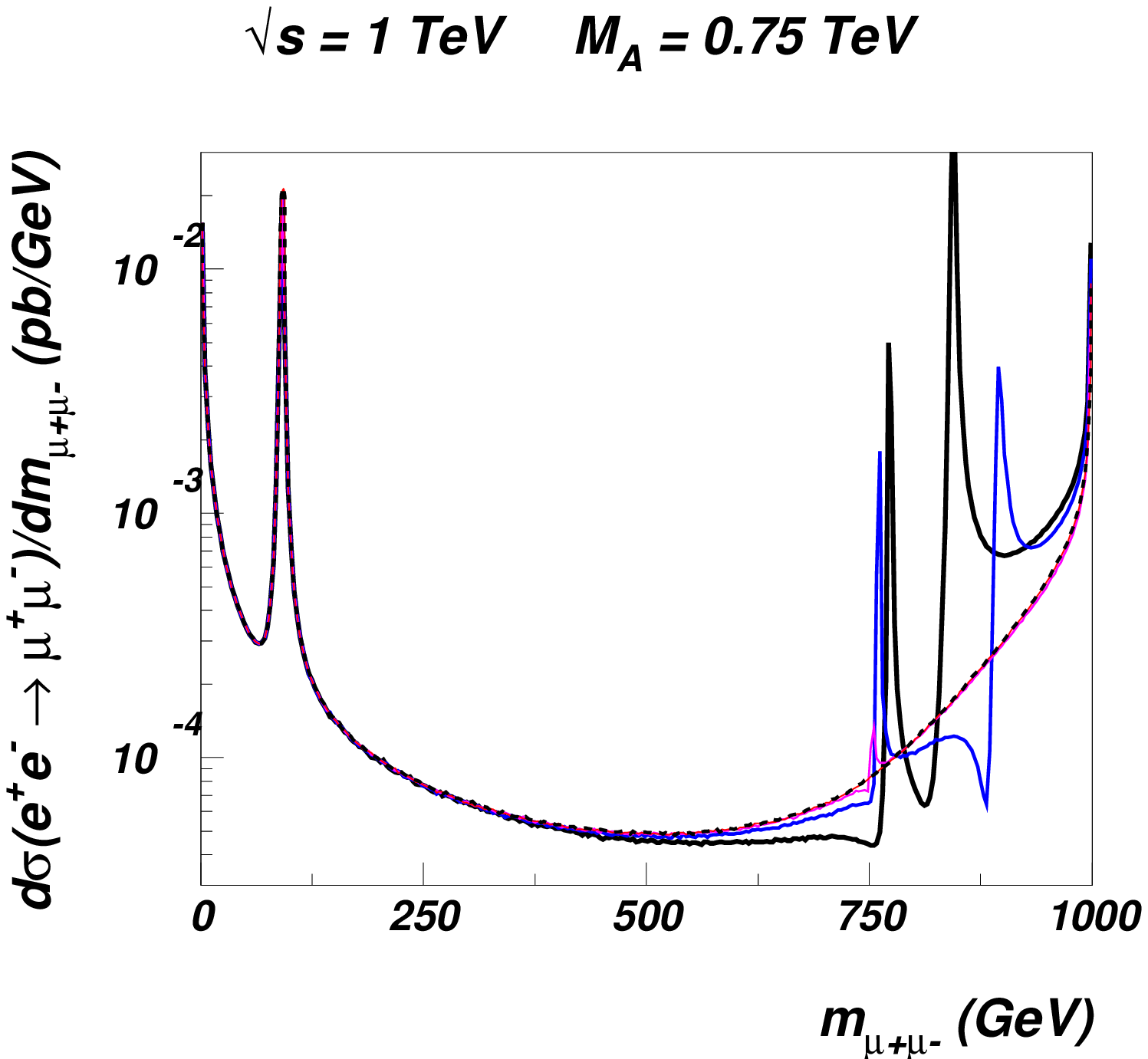}%
\includegraphics[width=0.45\textwidth]{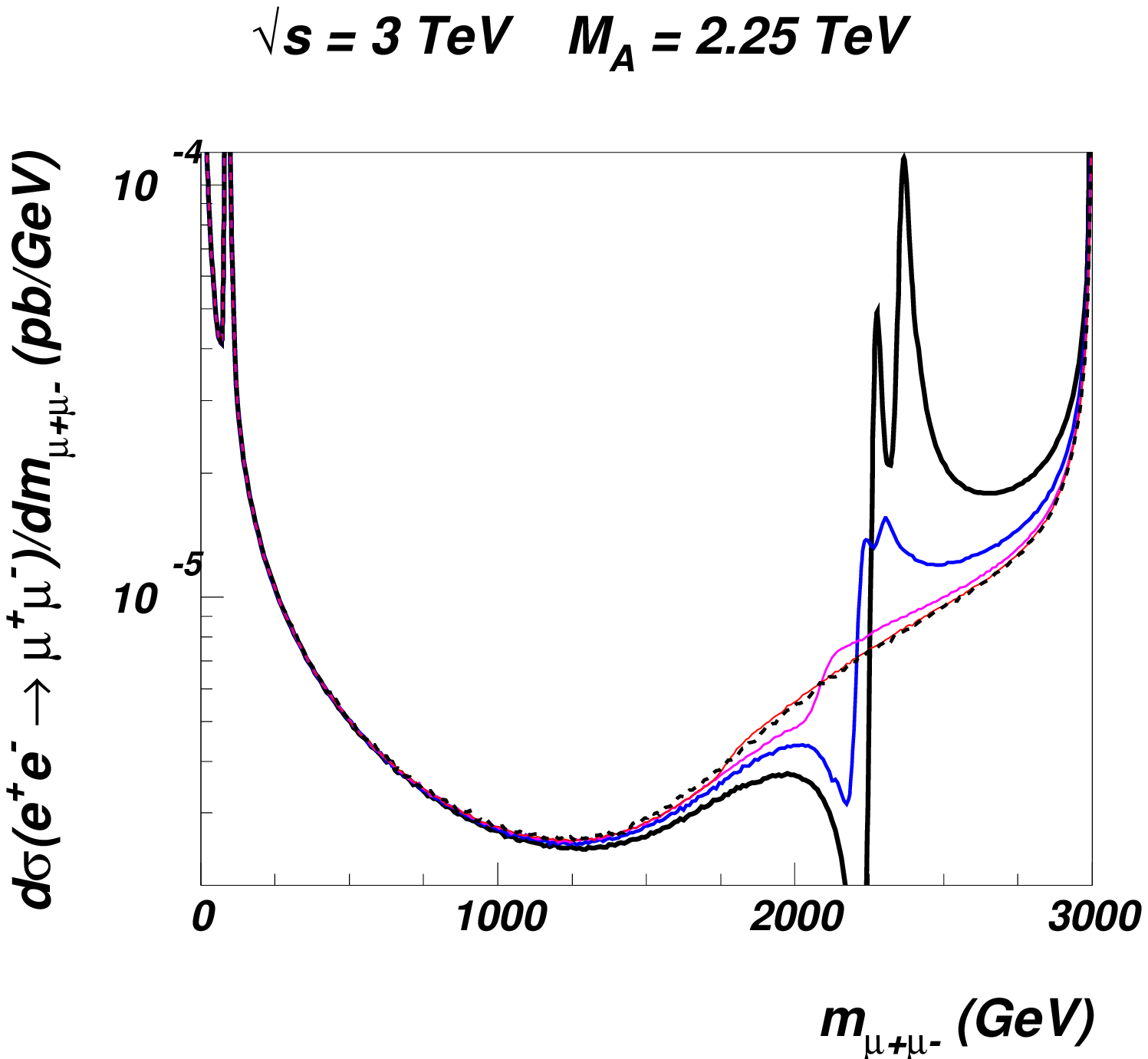}%
\caption{\label{fig:dimuon}Top 
row: Total cross-section for $e^+ e^-\to \mu^+\mu^-$  as a function of $\sqrt{s}$ with $M_A=0.75$~TeV (top left) and $M_A=2.25$~TeV (top right). Bottom row:  The differential cross section for $e^+ e^-\to \mu^+\mu^-$ as a function of the dimuon invariant mass. The thicknesses and colors of the solid lines indicate values of $\tilde g$. The used values are $\tilde g =2,3,5$, and 8 form the thick black lines to thin red ones. The dashed black line is the SM prediction. }
\end{figure}

\subsection{Dimuon and $WW$ final states}

The $e^+e^-\to \mu^+\mu^-$ cross section is shown in Fig.~\ref{fig:dimuon} (top) as a function of $\sqrt{s}$ at fixed $M_A$. The SM cross section is shown as dashed black line.
Peaks in the cross section are observed when the center-of-mass energy hits a resonance mass, tabulated in Table~\ref{table:masses}. As explained earlier, the peaks get suppressed with increasing $\tilde g$.

We also plot the dimuon differential cross section as function of the invariant dimuon mass in Fig.~\ref{fig:dimuon} (bottom). It is the initial state radiation which allows for a dependence on the dimuon mass. The composite states $R_1^0$ and $R_2^0$ are clearly identified for $\tilde g=2,3$, whereas for $\tilde g=5,8$ the peaks are hardly visible. The trend mimics the one observed when investigating the Drell-Yan production at the LHC  \cite{Foadi:2007ue}.

\begin{figure}[htb!]
\includegraphics[width=0.45\textwidth]{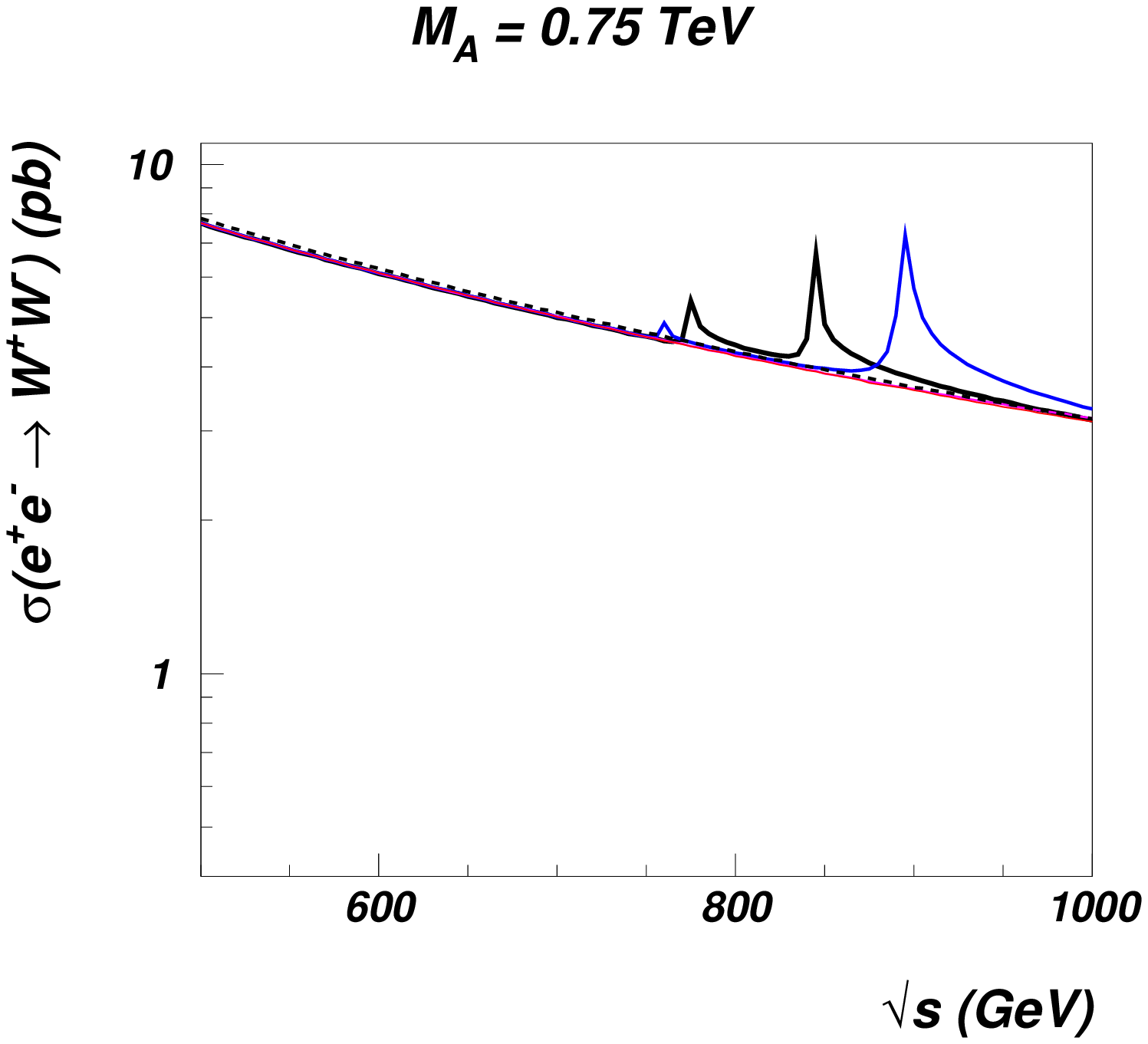}%
\includegraphics[width=0.45\textwidth]{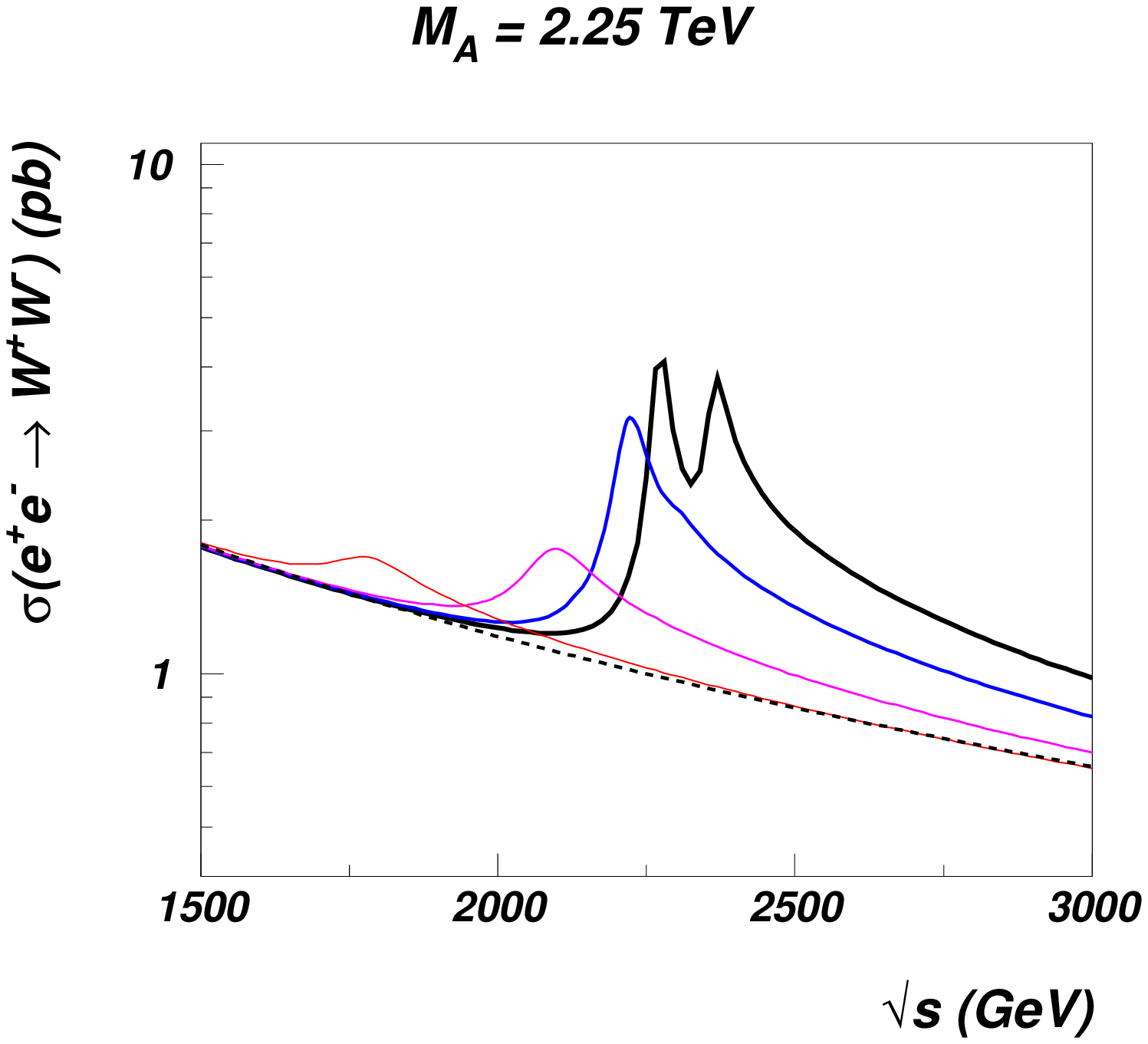}
\includegraphics[width=0.45\textwidth]{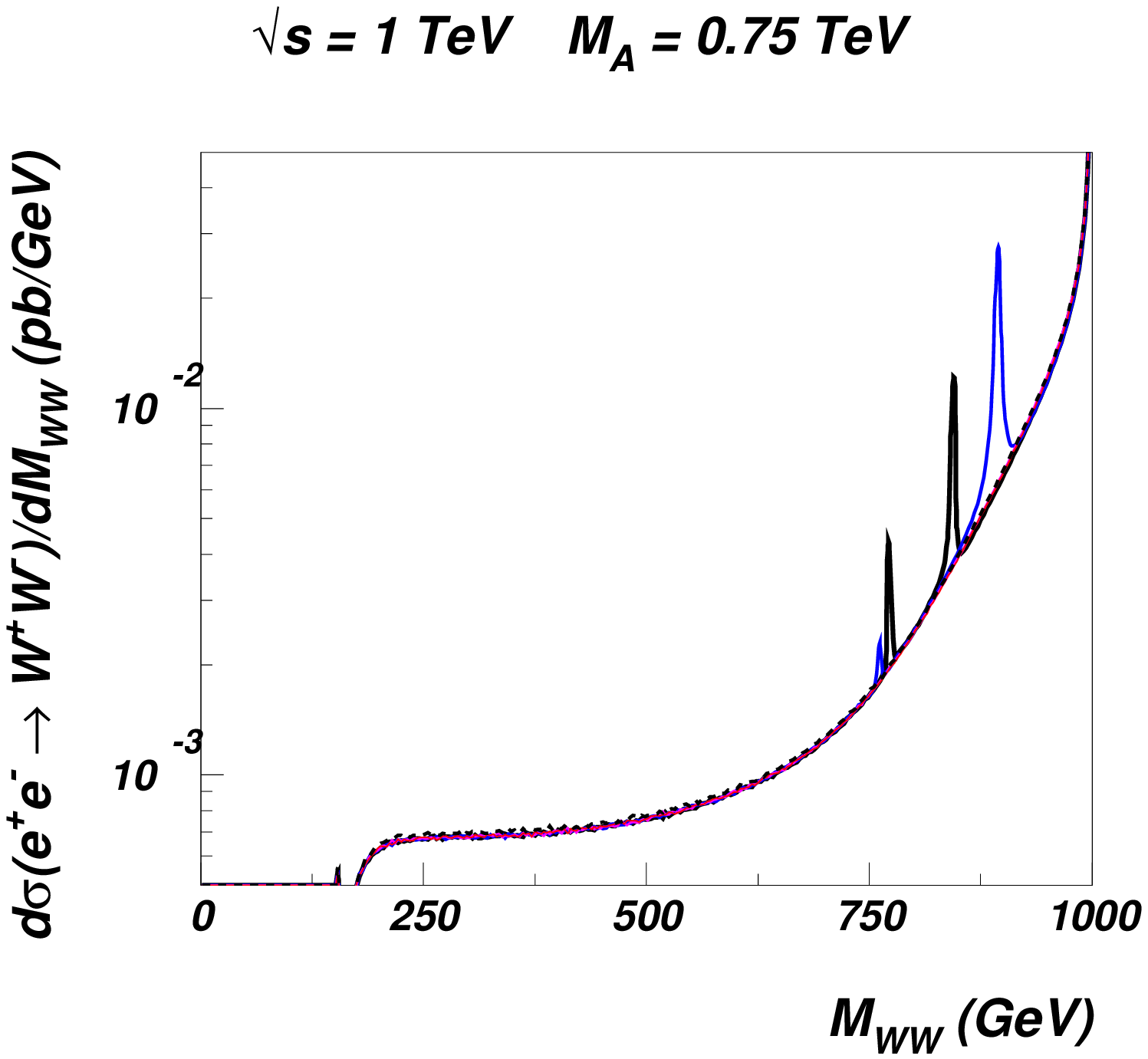}%
\includegraphics[width=0.45\textwidth]{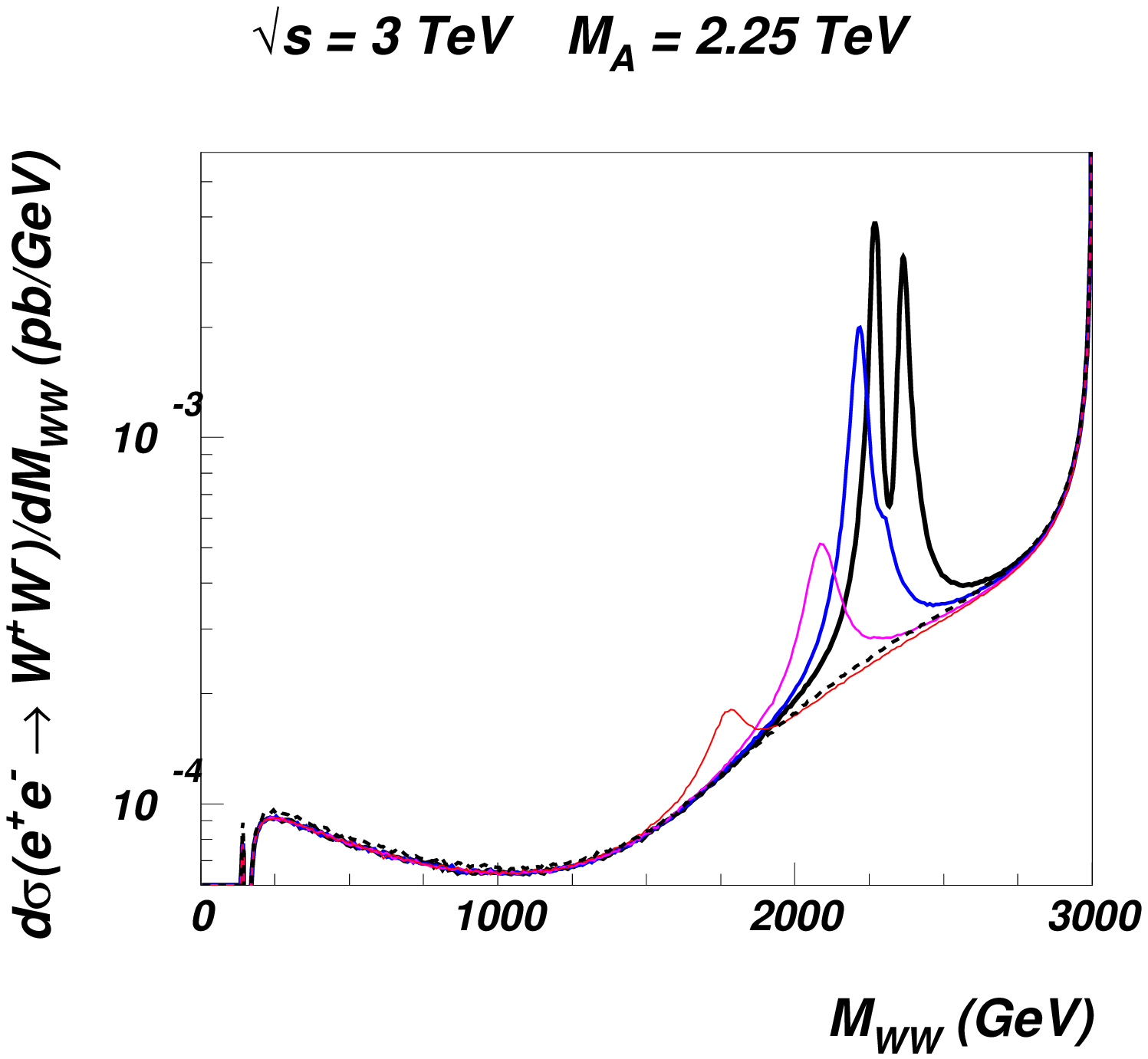}%
\caption{\label{fig:wwevents}Top row: Total cross-section for $e^+ e^-\to W^+W^-$  as a function of $\sqrt{s}$ with $M_A=0.75$~TeV (top left) and $M_A=2.25$~TeV (top right). Bottom row: Differential cross section for $e^+ e^-\to W^+ W^-$ as a function of the $M_{W^+ W^-}$ invariant mass.}
\end{figure}

The cross section for the $e^+e^-\to W^+W^-$ process as a function of $\sqrt{s}$ at fixed $M_A$ is shown in the top figures of Fig.~\ref{fig:wwevents}, while the differential cross section as function of the invariant mass is shown in the bottom plots of the same figure.  The cross section for the signal here is larger than the dimuon one for $\tilde g =5, 8$. This requires also $M_A$ to be large.

\subsection{Reach in dilepton and WW production}
\begin{figure}[htb!]
\includegraphics[width=0.45\textwidth]{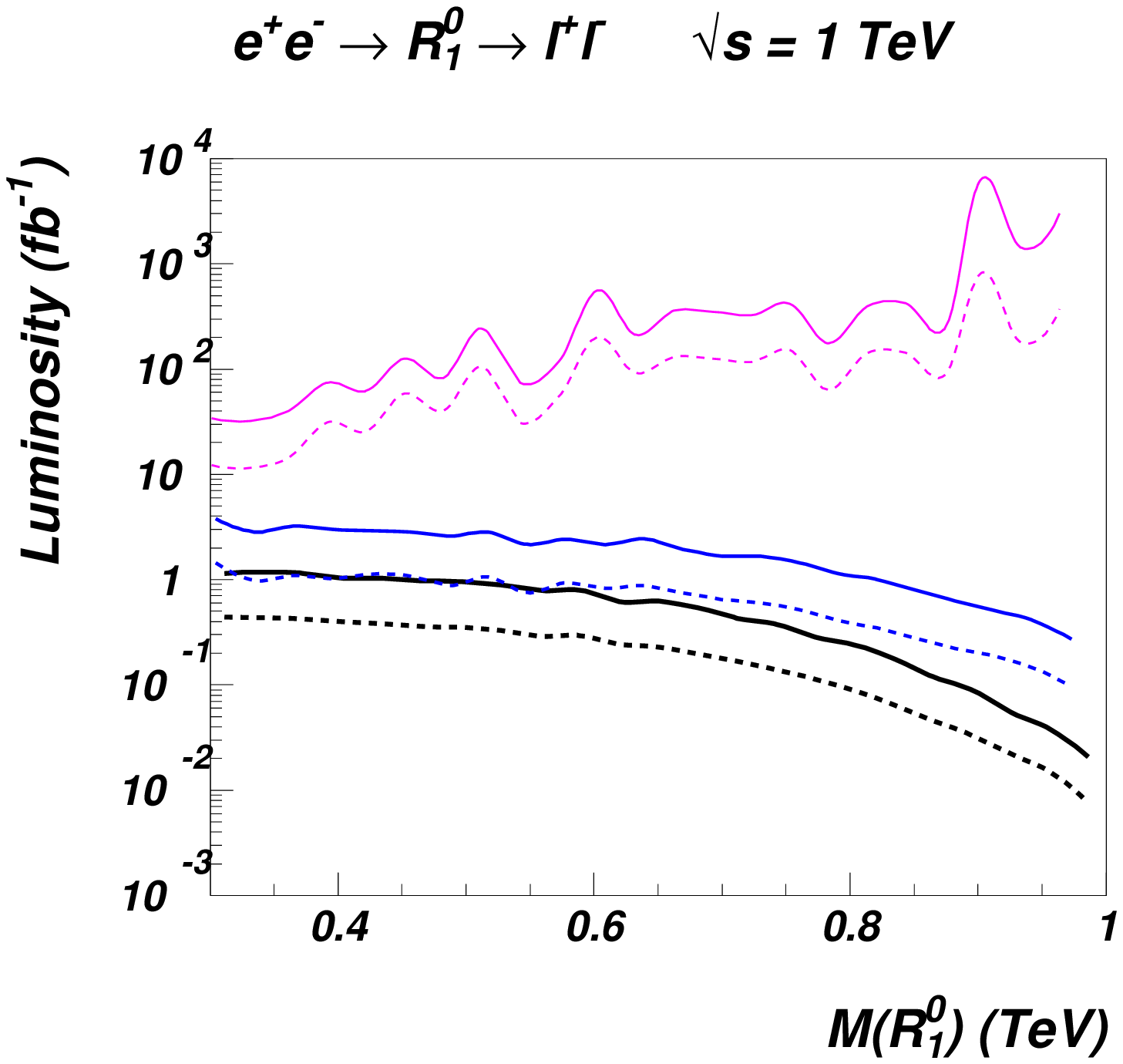}%
\includegraphics[width=0.45\textwidth]{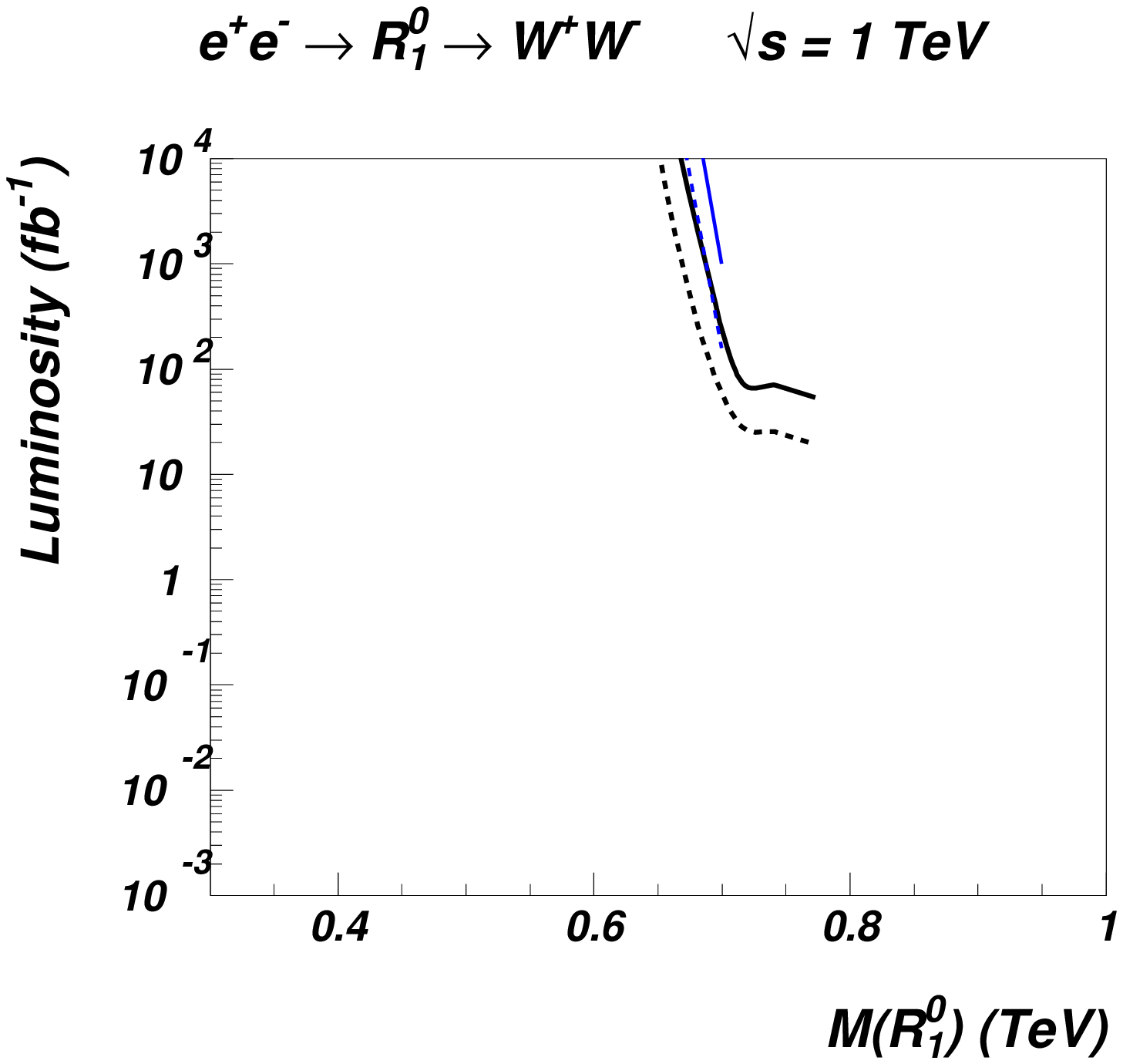}
\includegraphics[width=0.45\textwidth]{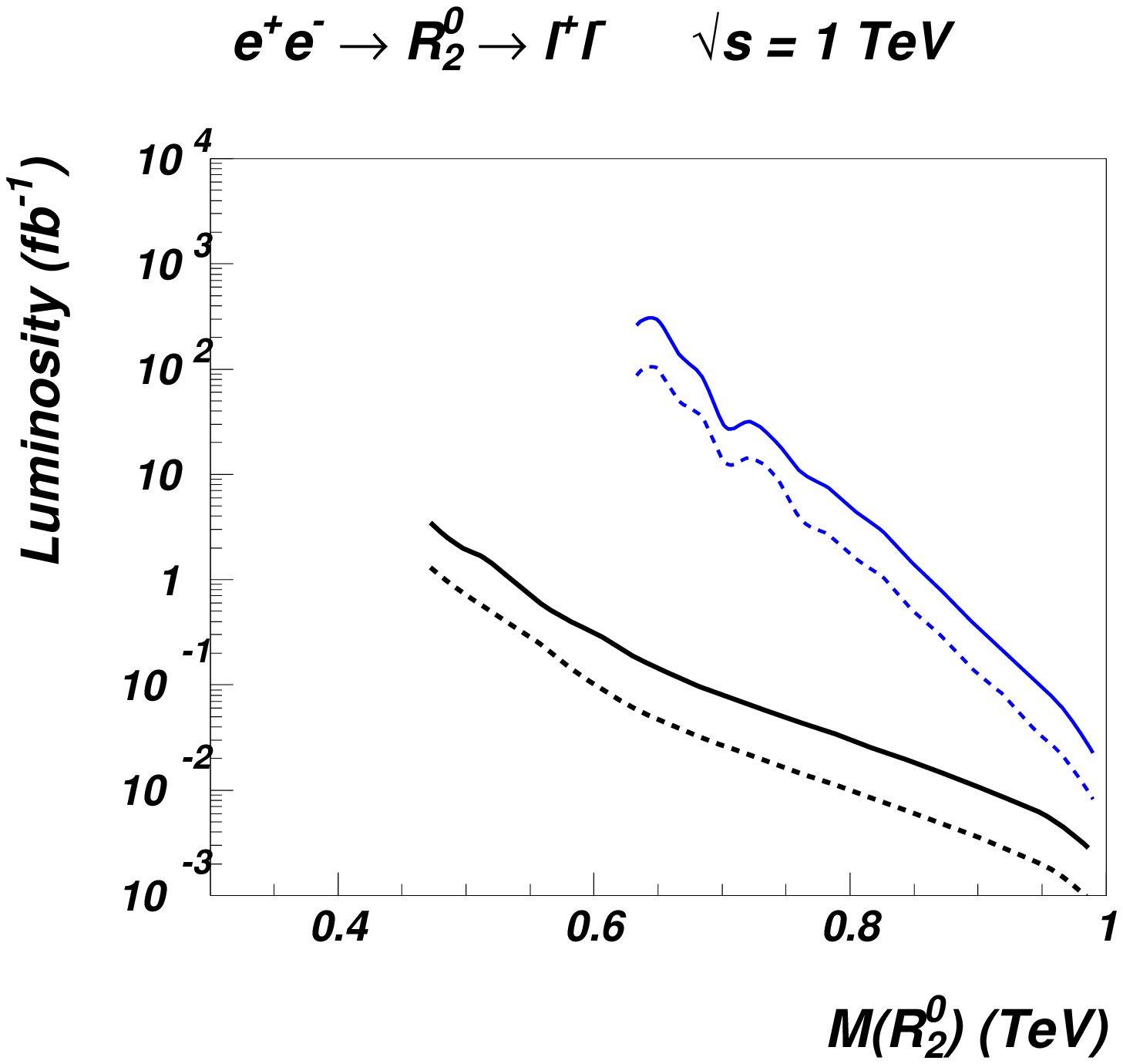}%
\includegraphics[width=0.45\textwidth]{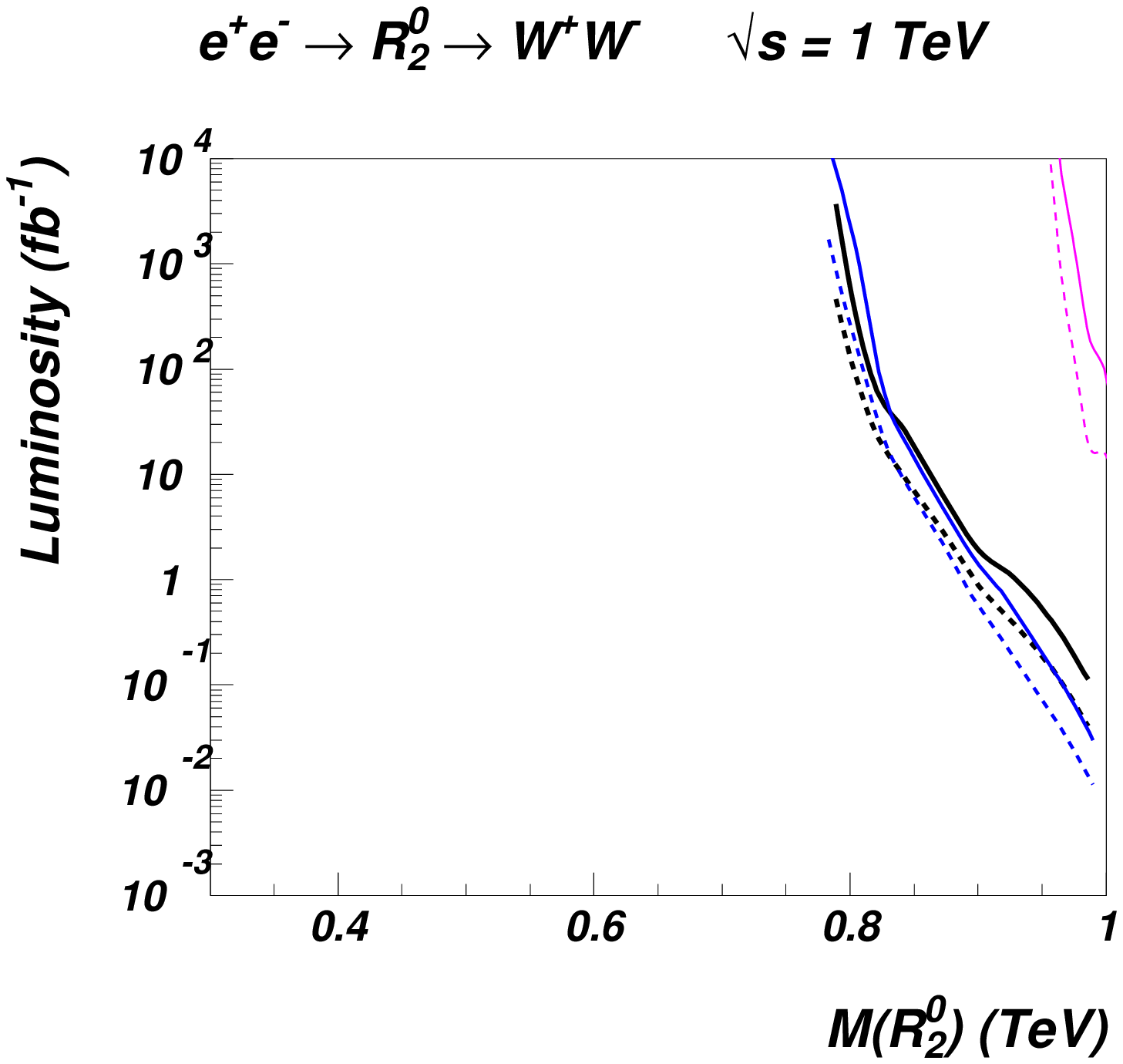}%
\caption{\label{lumiss1} Estimates for the required luminosity for $3\sigma$ (dashed lines) and $5\sigma$ (solid lines) discoveries of the vanilla technicolor vector resonances in dilepton and $WW$ production with $\sqrt{s} = 1$~TeV. The various lines are for $\tilde g =2,3,$ and 5 form  thick black lines to thin magenta lines.}
\end{figure}
\begin{figure}[htb!]
\includegraphics[width=0.45\textwidth]{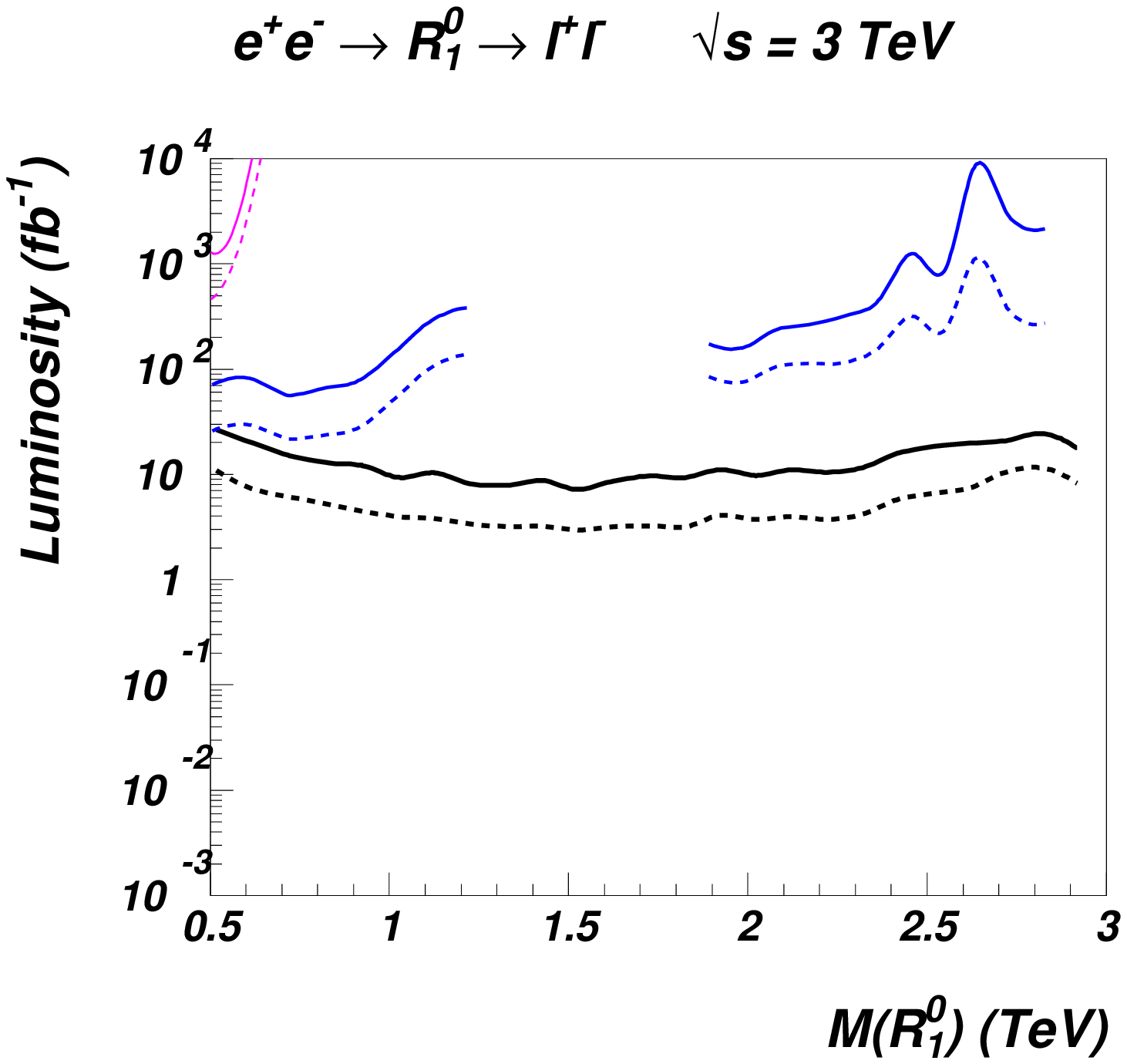}%
\includegraphics[width=0.45\textwidth]{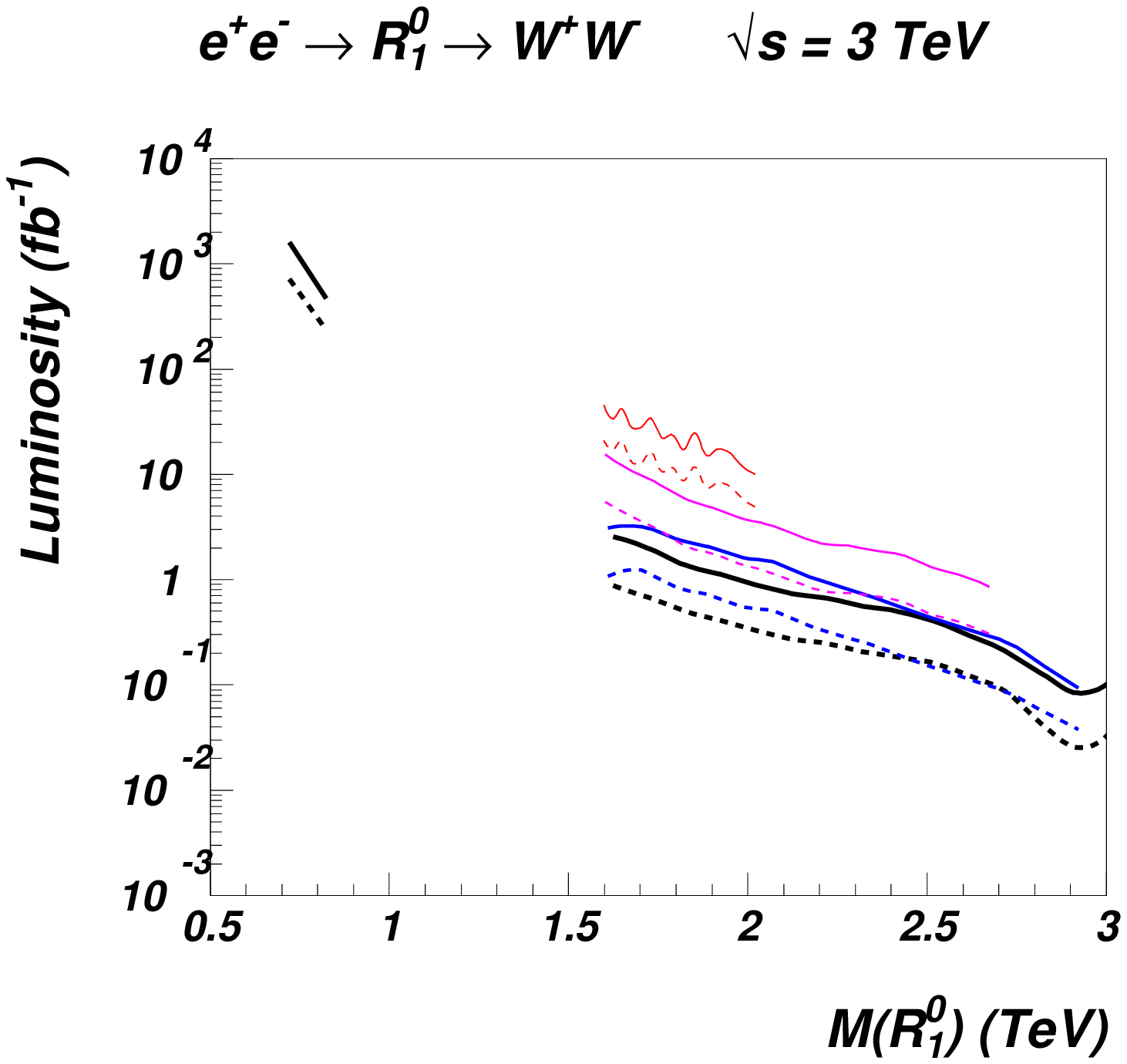}
\includegraphics[width=0.45\textwidth]{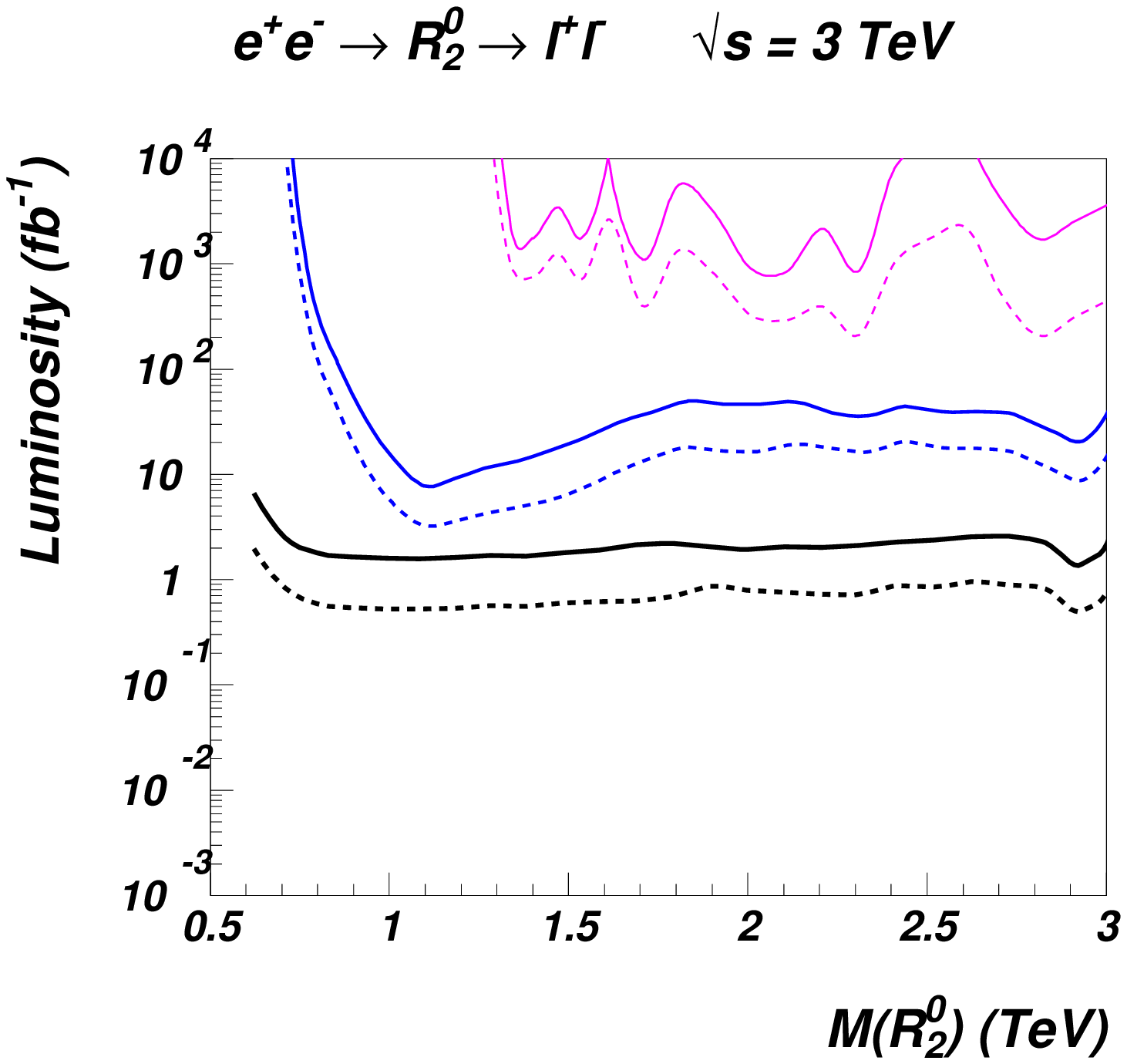}%
\includegraphics[width=0.45\textwidth]{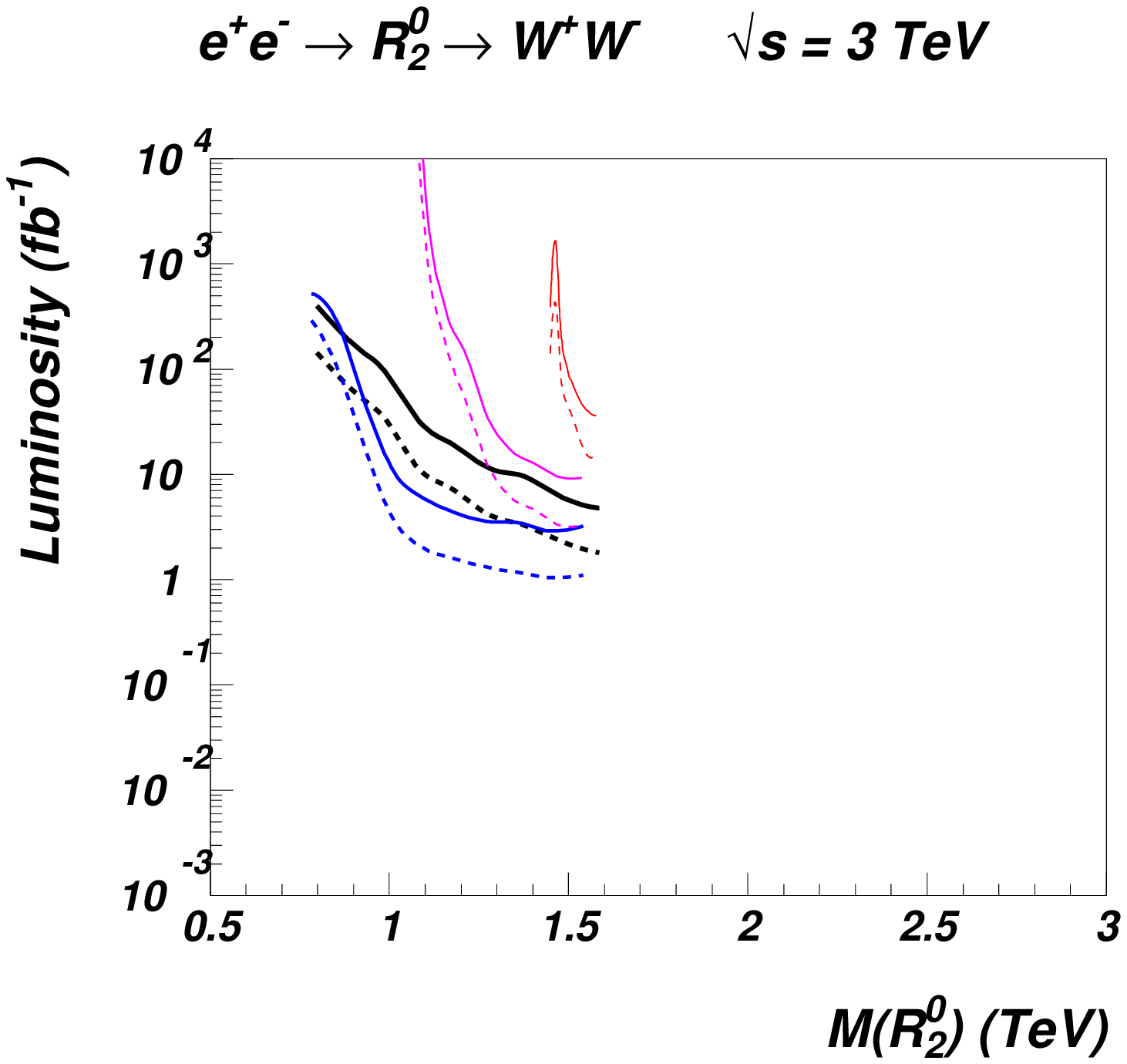}%
\caption{\label{lumiss3} The same as Fig.~\ref{lumiss1} but for higher  $\sqrt{s} = 3$~TeV. The various lines are for $\tilde g =2,3,5$, and 8 form thick black lines to thin red lines.}
\end{figure}

Let us now estimate the  luminosity needed to discover heavy spin-one states at LCs. We start with signature (1), i.e. $e^+e^- \to \ell^+\ell^-$. We use the following cuts:
\begin{eqnarray} \label{eq:acccuts}
p_\perp^\ell &>& 10~{\rm GeV},\qquad |\cos\theta^\ell| < 0.95 
\end{eqnarray}
where $\theta^\ell$ is the angle with respect to the beam direction. The precision of the energy reconstruction is simulated by adding a Gaussian random noise with variance $(0.15 GeV)^2 E/GeV$ to the energy of each lepton.
In addition we perform the following cut for the mass distribution near the resonance peak \cite{Basso:2009hf}
\beq \label{eq:peakcut}
 |M_{\ell\ell}-M_{R}| < \max\left(0.5\Gamma_R,0.15 \sqrt{\frac{M_R}{\rm GeV}} {\rm GeV}\right)
\eeq
where $R$ is either of the vector resonances. The two different cuts inside the $\max$ function come respectively from the width of the resonance and  the lepton energy resolution. $M_R$ and $\Gamma_R$ are calculated from the effective theory and shown in Table~\ref{table:masses}.

\begin{figure}[htb!]
\includegraphics[width=0.45\textwidth]{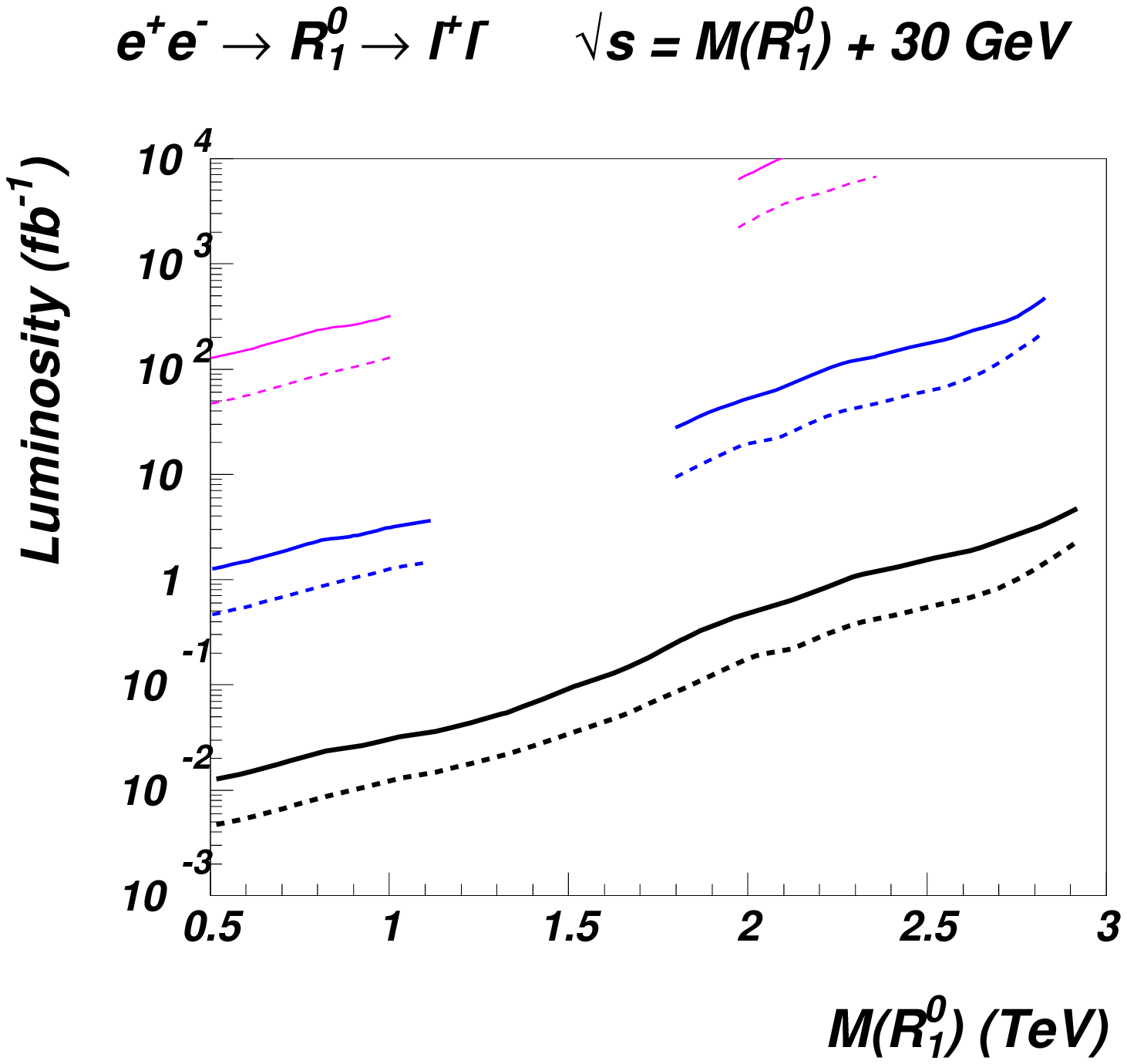}%
\includegraphics[width=0.45\textwidth]{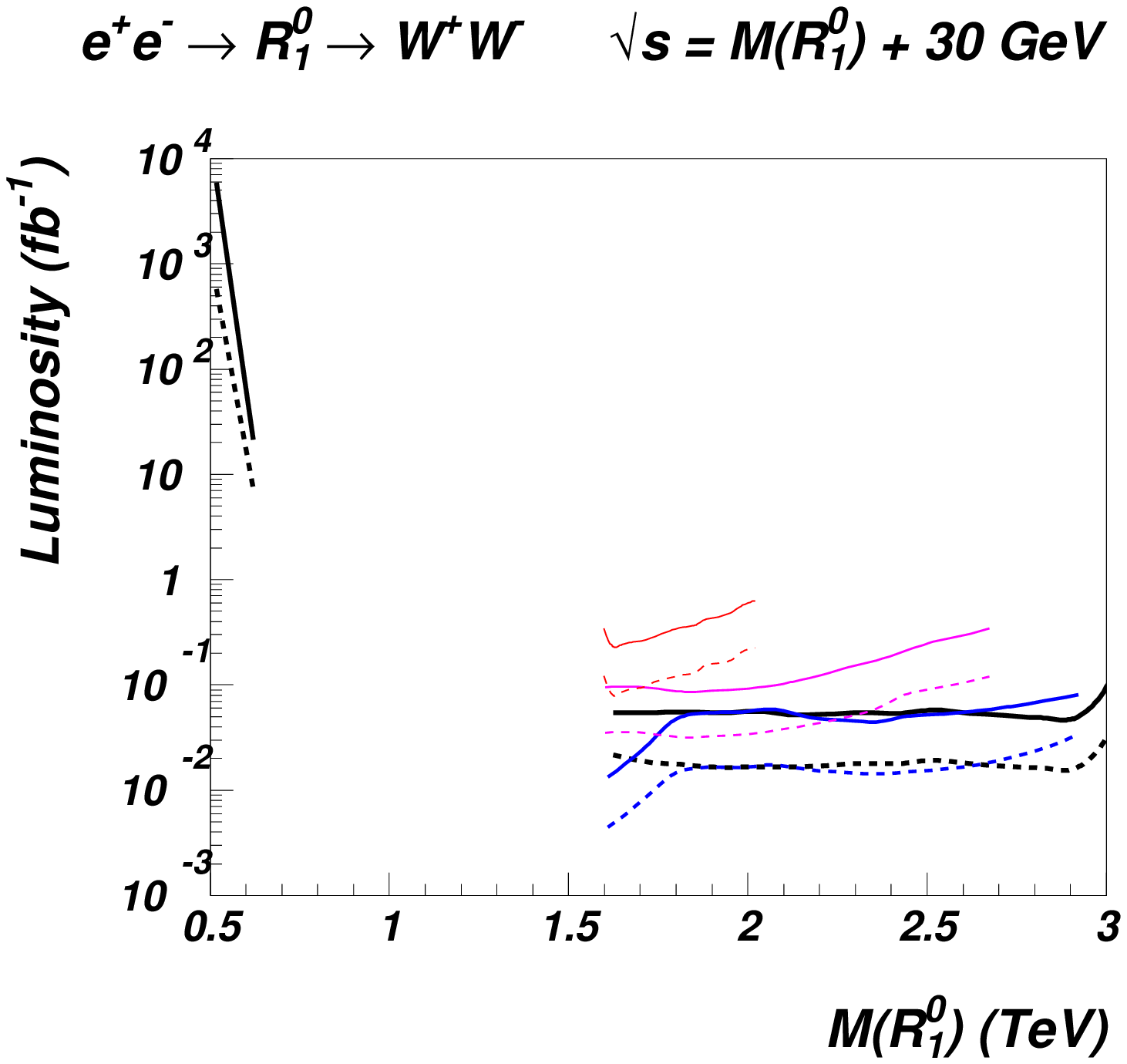}
\includegraphics[width=0.45\textwidth]{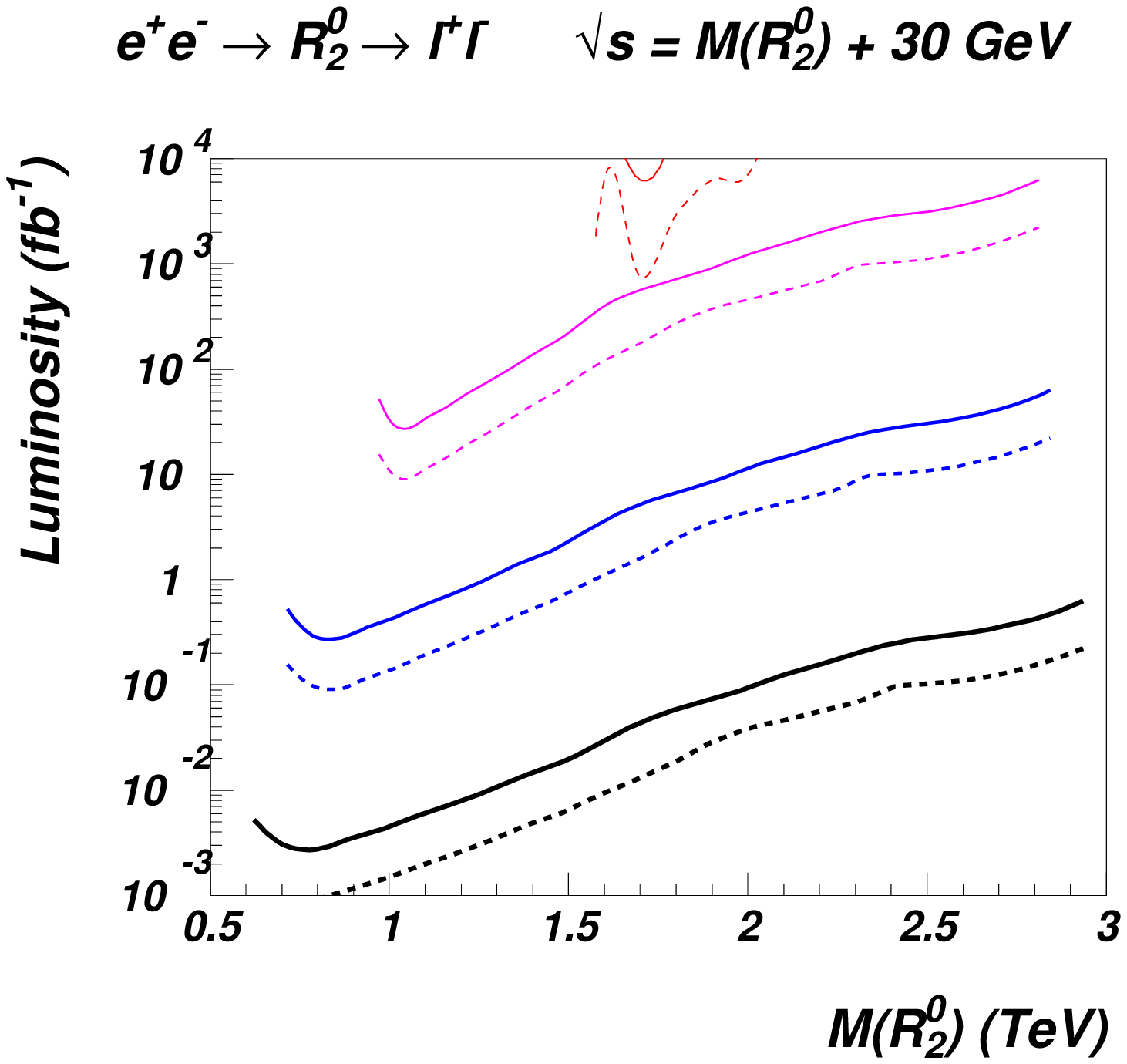}%
\includegraphics[width=0.45\textwidth]{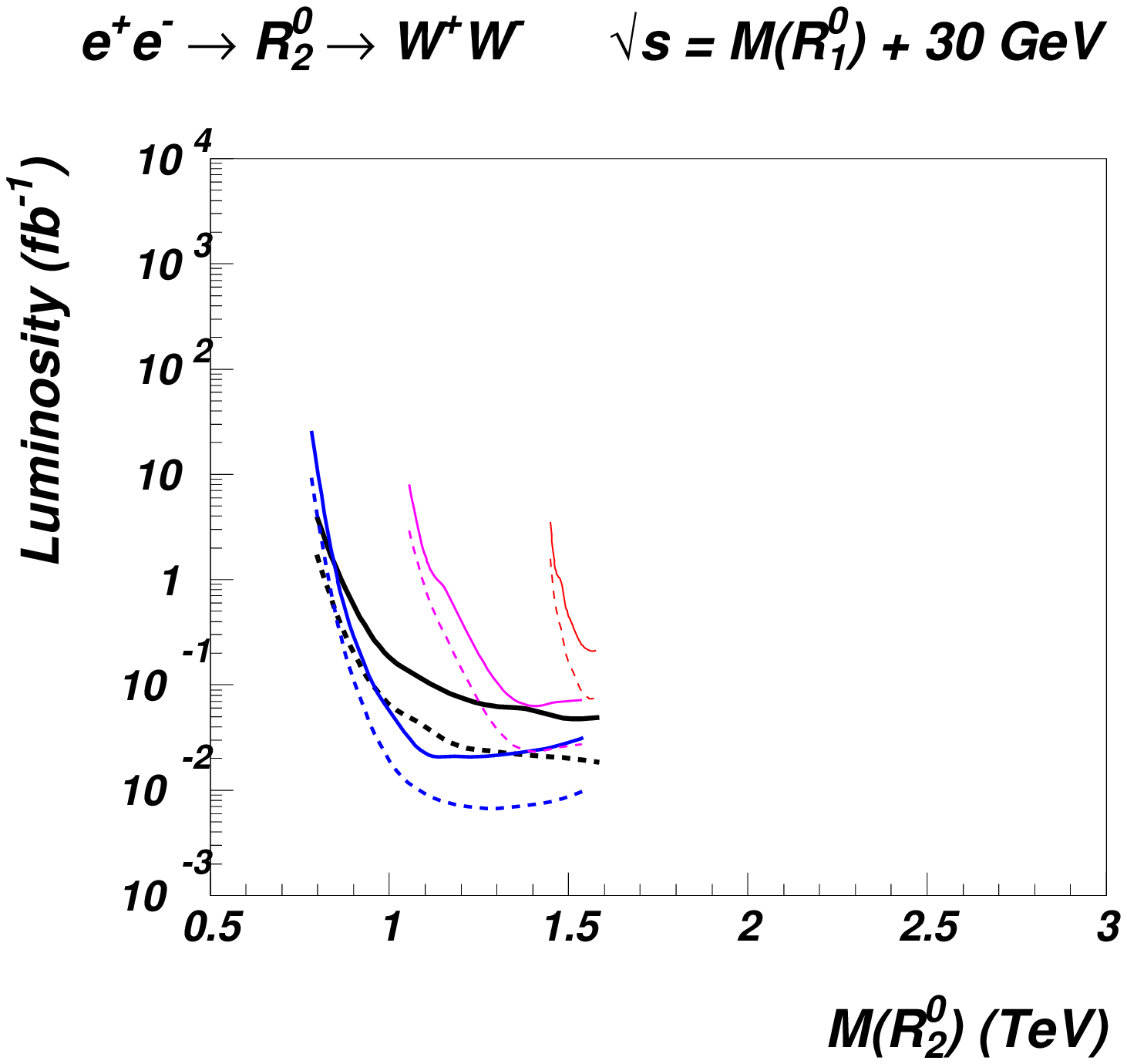}
\caption{\label{lumissvar} Estimates for the required luminosity for $3\sigma$ (dashed lines) and $5\sigma$ (solid lines) discoveries of the vanilla technicolor vector resonances with $\sqrt{s} = M_R+30$~GeV and with different values of the input parameters. 
}
\end{figure}

To a good approximation the background can be taken to be the SM prediction for this process. 
The signal is then defined as the excess of the vanilla technicolor over the SM background.
The significance of the signal is then defined as the number of
signal events divided by the square root of the number of
background events when the number of events is large, while a
Poisson distribution is used when the number of events is small.

We have also analyzed the signature (2),  
$W^+W^- \to \ell + \nu + 2 j$, by studying the distribution of the transverse mass variable
\begin{eqnarray}
  (M^{T}_{jj\ell})^2 &=&
  \left[\sqrt{M^2(jj\ell)+p_T^2(jj\ell)}
     +|\, \mpt|\right]^2-|\vec{p}_T^{}(jj\ell)+\mptv|^2 \ .
\end{eqnarray}
The jet energy reconstruction is simulated by smearing the energy $E$ of each jet with a Gaussian random error having variance $(0.5 GeV)^2 E/GeV$. The lepton energies are smeared as explained above for the dilepton final state. As the background, we use again the SM prediction for the same process. We cut the signal and background as follows. We require that both leptons and jets have $|\cos\theta|<0.95$ with respect to the beam axis, the lepton has $|p_\perp|> 10$~GeV and that the jets have $|p_\perp|>20$~GeV. We add the cut 
\begin{equation}
 |M^T_{jj\ell}-M_{R}| < \max\left(0.5\Gamma_R,0.5 \sqrt{\frac{M_R}{\rm GeV}} {\rm GeV}\right) \ ,
\end{equation}
where the latter expression inside $\max$ estimates the jet energy resolution which is the dominating error source. 
Notice that, in particular for high $M_R$ the $W$ decaying to two jets will be highly boosted, and hence the jets easily merge. However, we did not require a jet separation in our analysis, so we effectively count the single jet events in the signal as well. This improves the signal since there is no additional single jet background. The significance is calculated as for the $\ell^+\ell^-$ signature above.

The calculated significances for $e^+e^- \to \ell^+\ell^-$ and for $e^-e^+ \to W^+W^-$ are used to estimate the necessary luminosities to discover the heavy vector states for various parameter values in Figs.~\ref{lumiss1},~\ref{lumiss3}, and~\ref{lumissvar}.

First, let us take the center-of-mass energy to be fixed to the maximum operating energy of a given LC. 
Figs.~\ref{lumiss1} and \ref{lumiss3} show the estimates for $3\sigma$ and $5\sigma$ discoveries with $\sqrt{s} = 1$~TeV and 3~TeV, respectively. We used the reference values $\tilde g=2,3,5,$ and 8 and scanned over $M_A$ from 0.5~TeV to 3~TeV, covering roughly the allowed parameter
space of Fig.~\ref{fig:bounds}.
In particular, for $\tilde g=3,5$ the two resonance peaks tend to overlap for some values of $M_A$. In such cases we only show the signal for the dominant resonance peak. This is why some of the lines end abruptly in the plots. For example, the $R_2$ resonance dominates the $W^+W^-$ signal for $\sqrt{s}=1$~TeV (right hande side plots in Fig.~\ref{lumiss1}), $\tilde{g}=2$ and when the resonance masses are near 1~TeV. Therefore our simple estimate for the $R_1$ reach fails, and is not shown. A more sophisticated analysis would be necessary to estimate the size of the weaker signal in the excluded regions.

As expected from the distribution plots of the previous section, the required luminosity for the $\ell^+\ell^-$ signature (left hand plots) increases strongly with $\tilde g$. 
However, the $\tilde g$ dependence of the $W^+W^-$ estimates (right hand plots) is weaker, and therefore the required luminosities are lower than for the dilepton final state for high values of $\tilde g$.

It is also interesting to study the capability of the collider to find resonances in the scanning mode (varying $\sqrt{s}$) and its ability to confirm weak signals observed at the LHC. The reach for this kind of situation is presented in
Fig.~\ref{lumissvar}, where the center-of-mass energy is tuned to $M_{R_i}+30$~GeV. The value of the shift $30$~GeV between the resonance mass and $\sqrt{s}$ was chosen to be slightly larger than a typical vector width in our model. Notice that for $\tilde g = 2$ one has that $R_2^0$ appears to be detectable at very low luminosities. 
For high $\tilde g$ and masses above 2~TeV the resonances are very broad, and our approach is expected to underestimate the signal. 
Notice the peculiar structure in the $W^+W^-$ plots (visible on right hand side in Figs.~\ref{lumiss3} and~\ref{lumissvar}): we have estimates for $R_1^0$ for large $M_{R_1}$, which end suddenly at $M_{R_1}=1.6$~TeV and continue almost smoothly on the $R_2^0$ plot below for $M_{R_2}<1.6$~TeV. This is due to the level crossing between the axial and vector spin-one states in our model, which occurs at $1.6$~TeV. As pointed out above, the $W^+W^-$ signal \cite{Foadi:2007ue} is dominated by the vector spin-one resonance, while the axial one mostly decays to $HZ$. The axial vector signal can be easily separated from the vector one only in the regime of very small $M_{R_1}$.

\begin{figure}[htb!]
\includegraphics[width=0.45\textwidth]{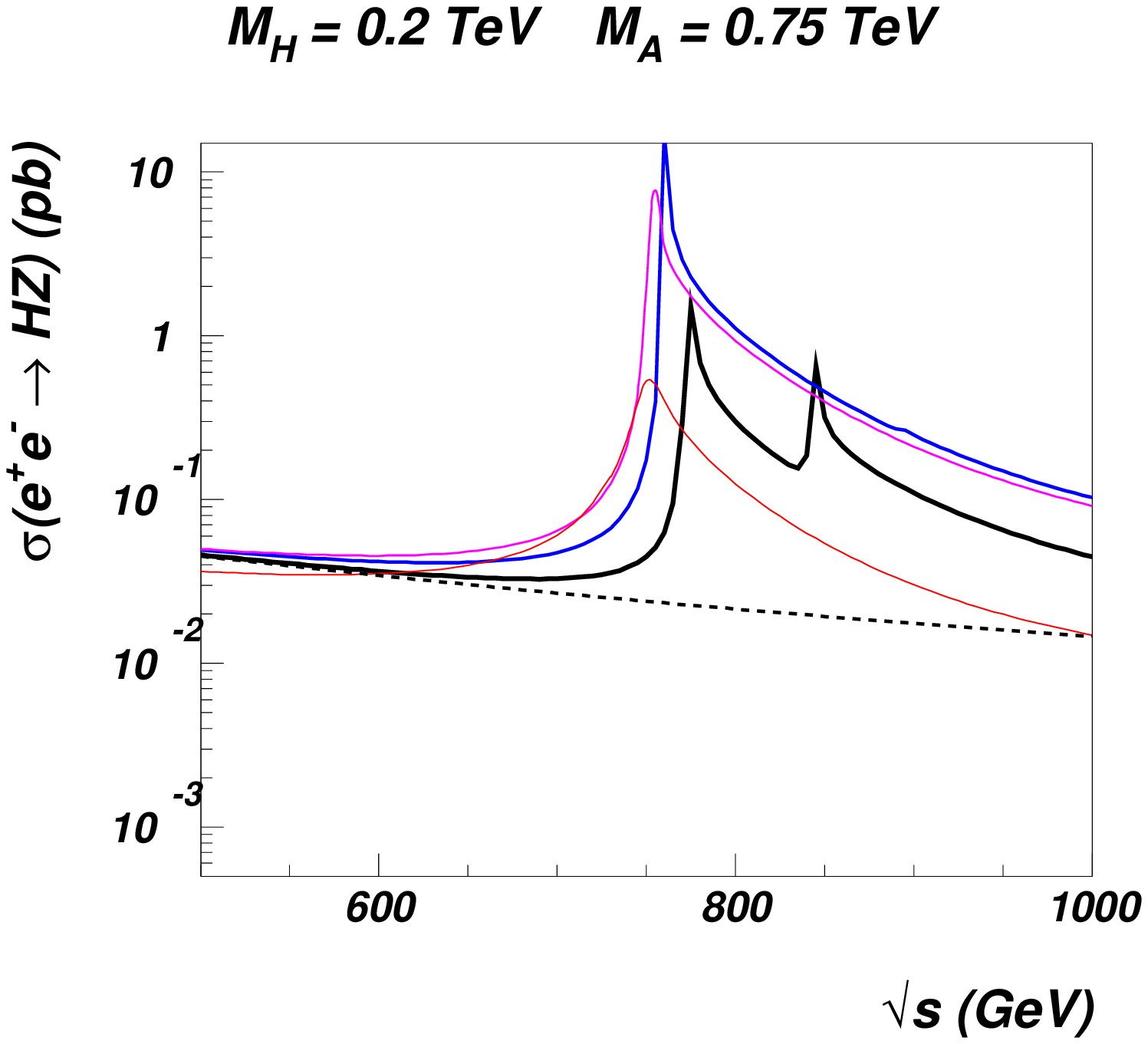}%
\includegraphics[width=0.45\textwidth]{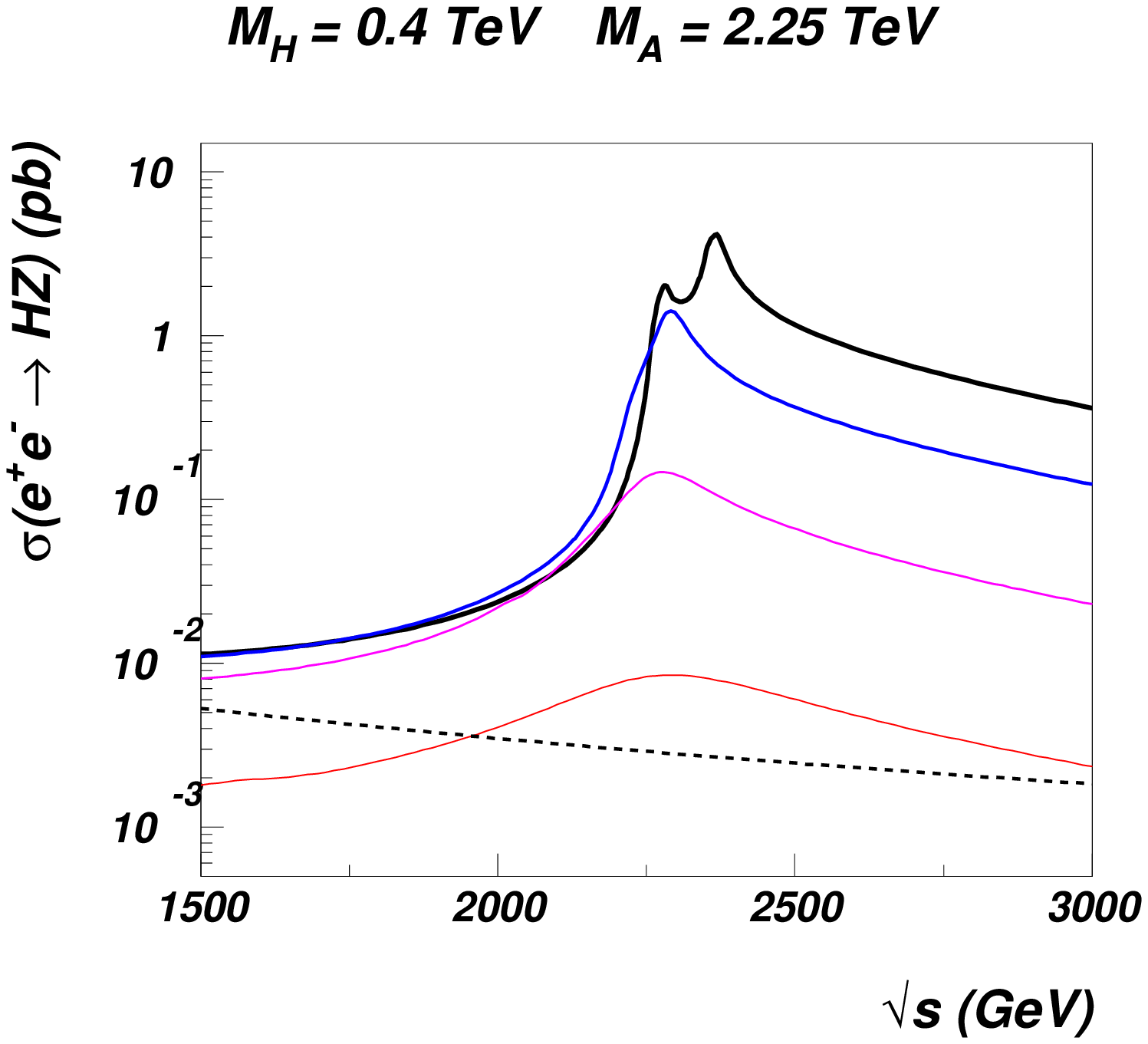}
\includegraphics[width=0.45\textwidth]{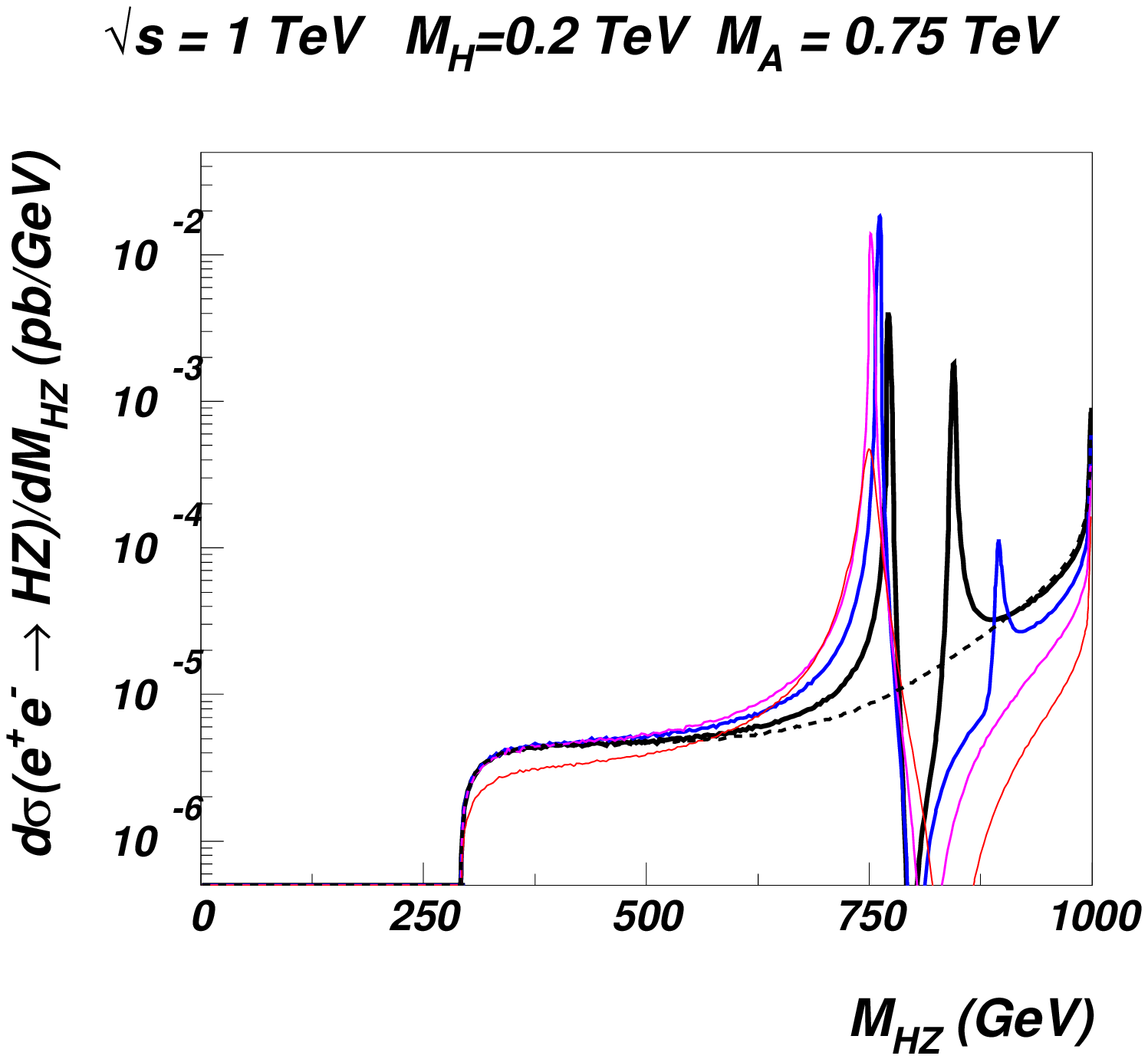}%
\includegraphics[width=0.45\textwidth]{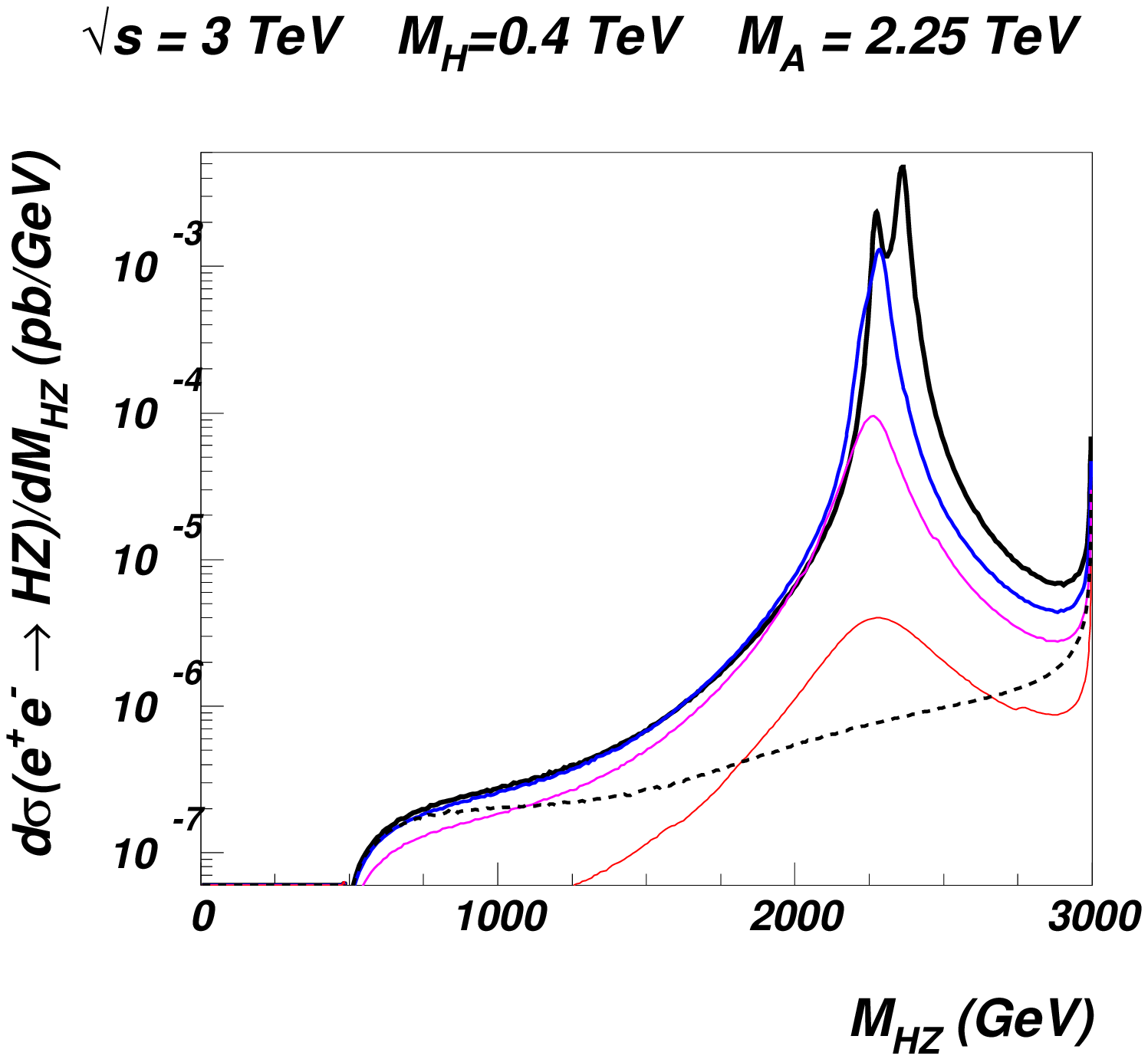}%
\caption{\label{eehzprod} Top row:  
The cross section as a function of $\sqrt{s}$ with $M_A=0.75$~TeV (top left) and $M_A=2.25$~TeV (top right). Bottom row: The differential cross section for $e^+ e^-\to ZH$ as a function of $M_{ZH}$. The various lines are for $\tilde g =2,3,5$, and 8 form thick black lines to thin red lines. }
\end{figure}

\subsection{6 particle final states from $HZ$ production}

The presence of the heavy vectors can lead to an enhancement of the composite Higgs production in association with a SM vector bosons, as pointed out in \cite{Zerwekh:2005wh,Belyaev:2008yj}. Since, in models of minimal walking technicolor, the composite Higgs is expected to be heavy with respect to $M_Z$ but light \cite{Hong:2004td,Dietrich:2005jn,Dietrich:2006cm,Sannino:2009za,Sannino:2008ha} with respect to the scaled up sigma in QCD \cite{Sannino:1995ik,Harada:1995dc,Harada:1996wr,Black:1998wt,Harada:2003em}, the 6 particle final states from the associate Higgs production is therefore an appealing discovery channel \cite{Belyaev:2008yj}. 

The Higgs production amplitude is proportional to the coupling strength of the Higgs to the vector states.
At the effective Lagrangian level this coupling is a free parameter with its value depending, at least, on the coupling $s$  \cite{Foadi:2008xj}. We shall use $s=0$ which still allows for a reasonable estimate of the expected order of magnitude for this process.

In Figure~\ref{eehzprod} we present the $HZ$ production
cross section, compared the SM background as a
function of the center-of-mass energy. 
The inclusion of the composite vectors leads to an highly enhanced cross section with respect to the SM one in most parts of the parameter space.
We also plot the differential cross section as a function of the $HZ$ mass in Fig.~\ref{eehzprod} (bottom). In contrast to the dimuon production the peaks associated to the heavy vector states $R_{1,2}^0$ are clearly visible also for $\tilde g=5, 8$.

\begin{figure}[htb!]
\includegraphics[width=0.45\textwidth]{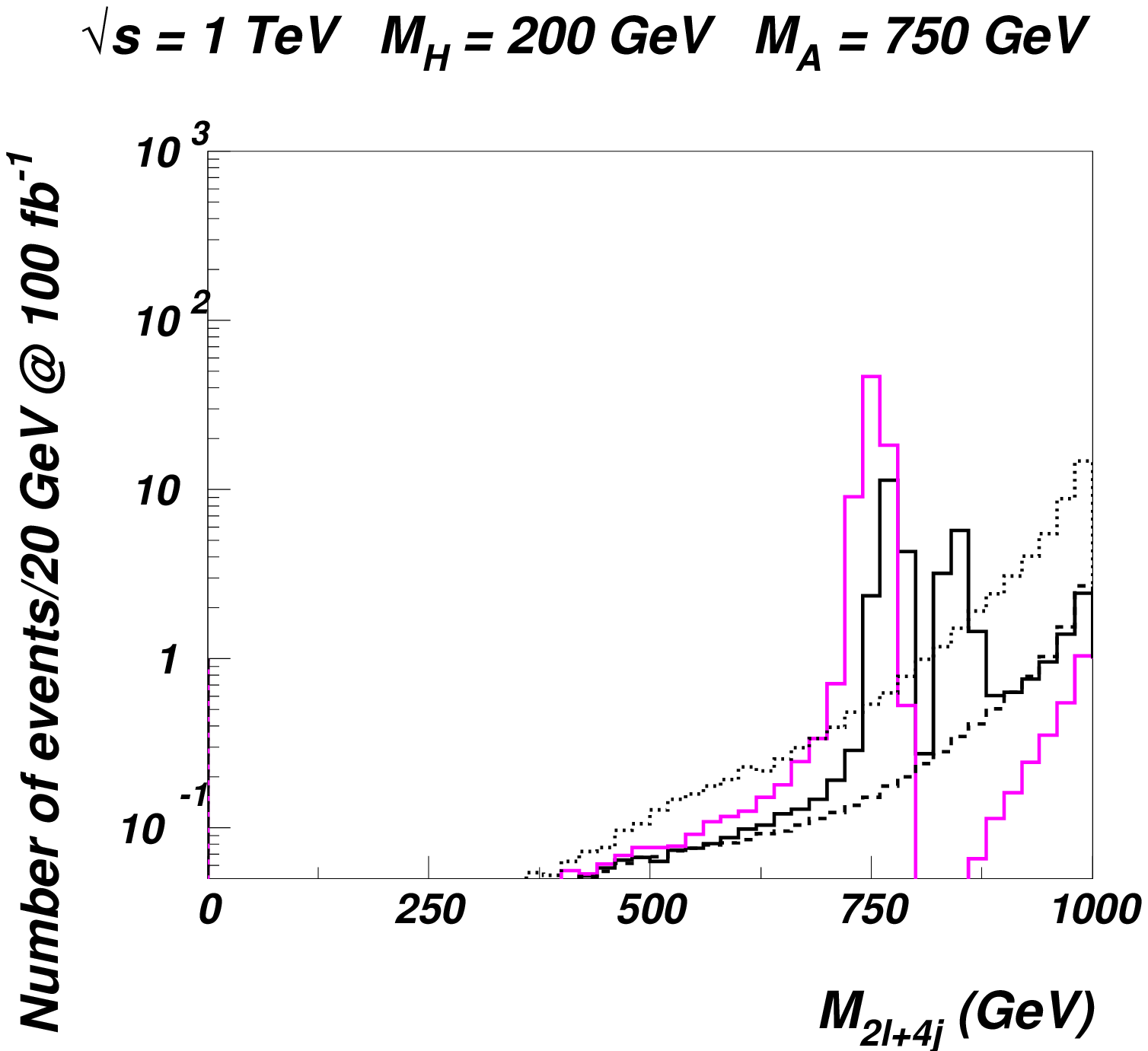}%
\includegraphics[width=0.45\textwidth]{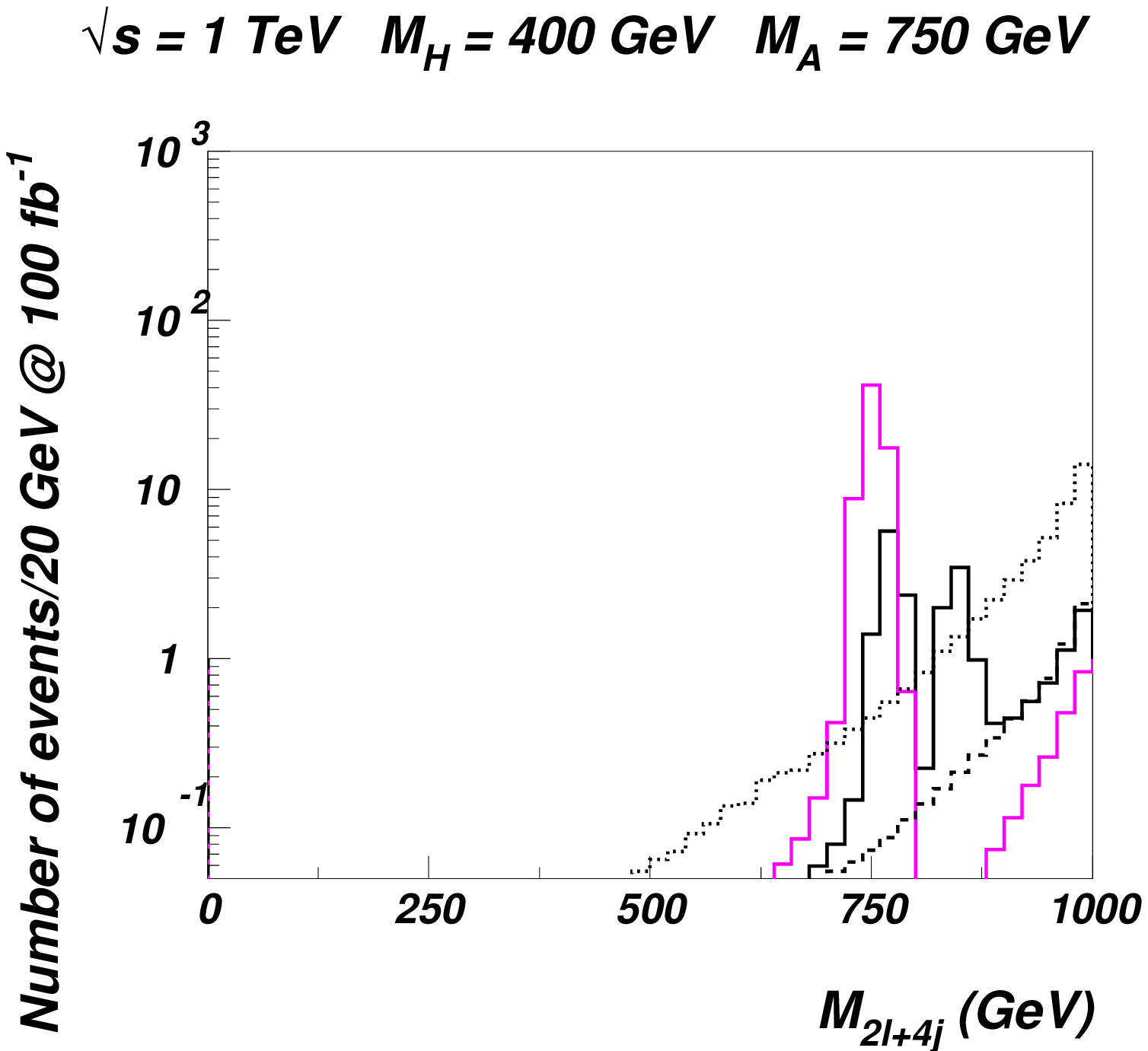}
\includegraphics[width=0.45\textwidth]{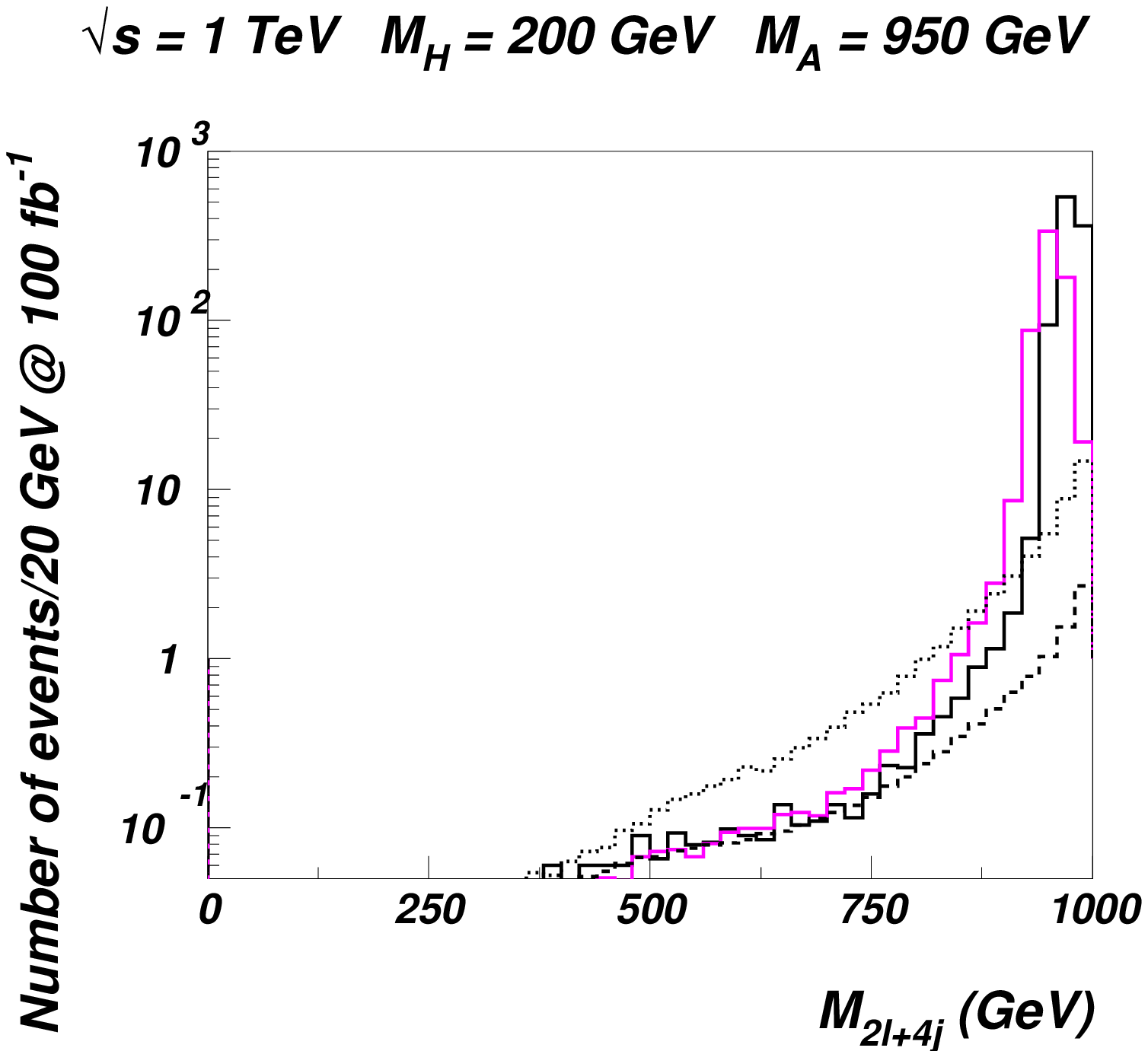}%
\includegraphics[width=0.45\textwidth]{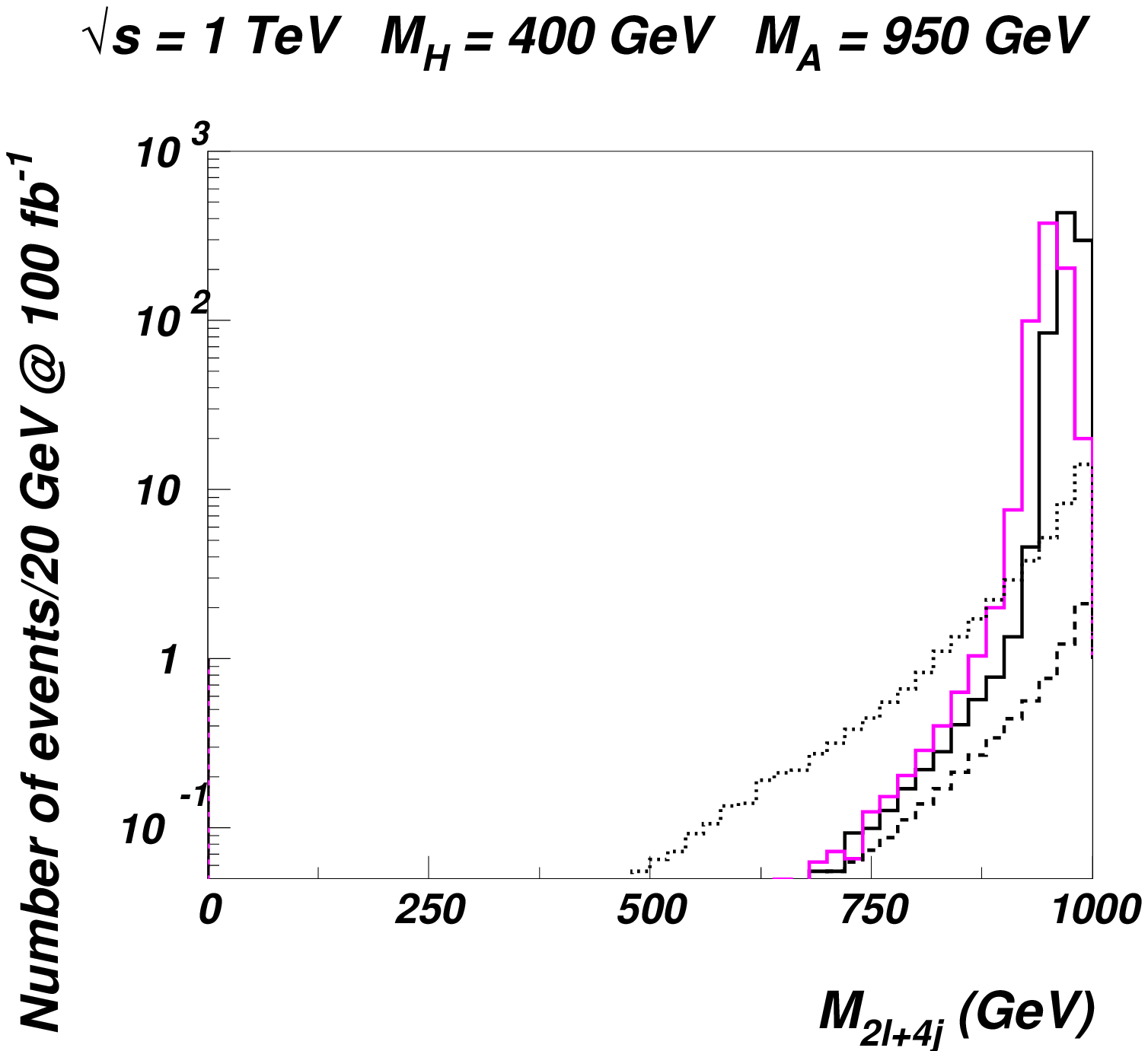}
\caption{\label{2l4jlm}Distribution of events for the $e^+e^- \to 2\ell+4j$ signature at $\sqrt{s}=1$~TeV and $L=100\ fb^{-1}$. Continuous lines are the distribution for signal events, and dotted lines are the background.   The value of  $\tilde{g}=2$ corresponds to the black solid line while the magenta corresponds to  $\tilde{g}=5$.}
\end{figure}
\begin{figure}[htb!]
\includegraphics[width=0.45\textwidth]{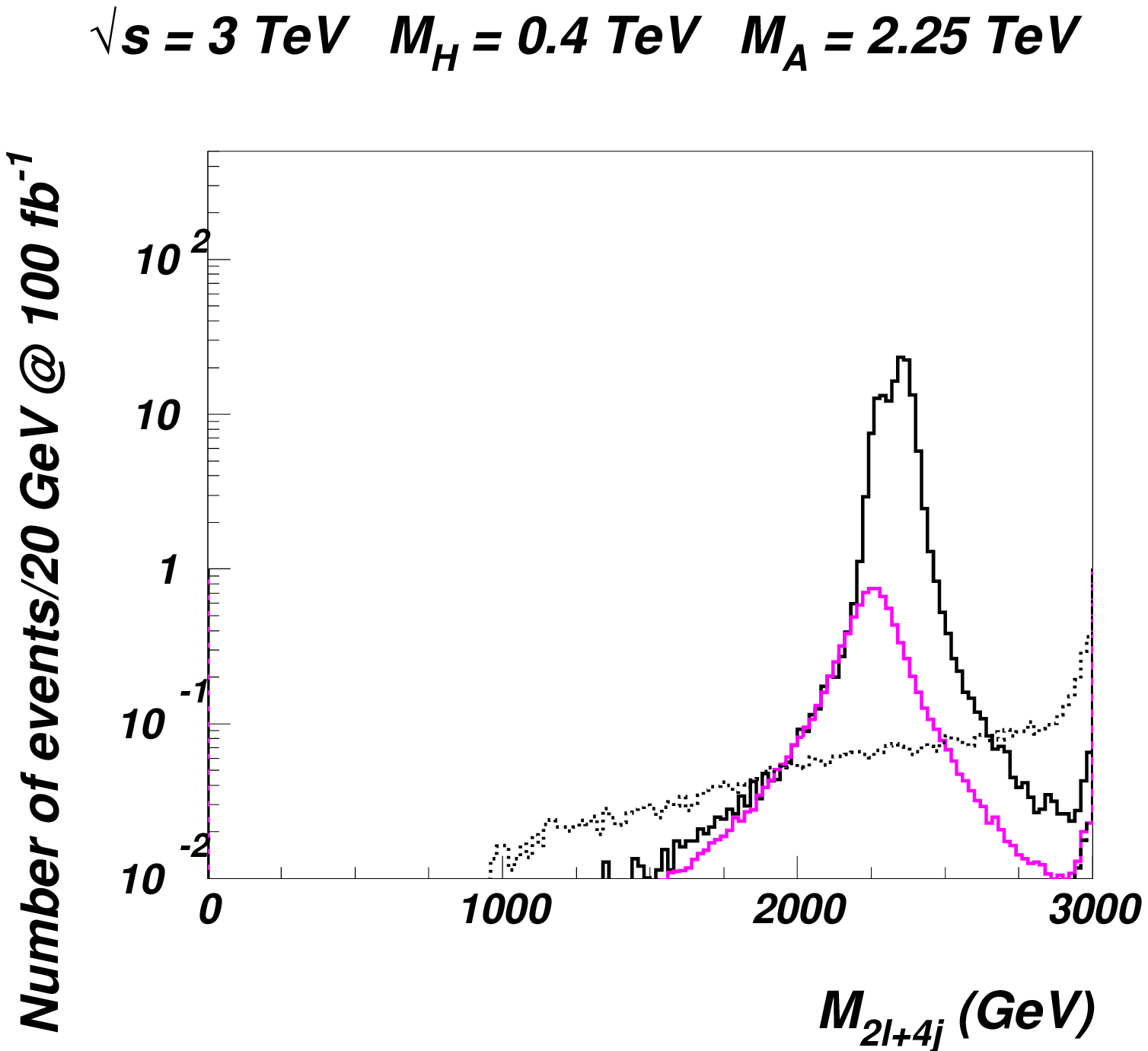}%
\includegraphics[width=0.45\textwidth]{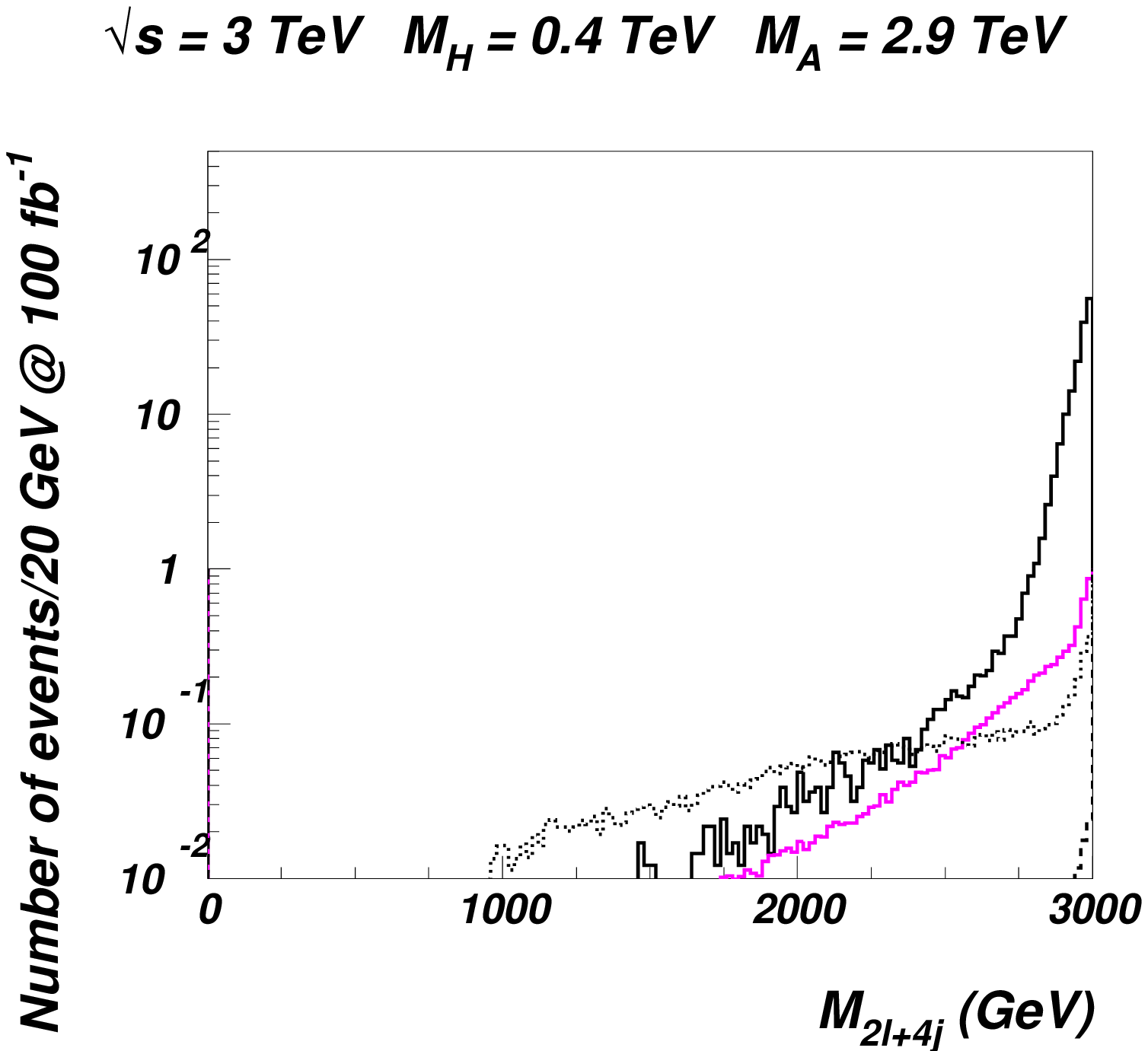}%
\caption{\label{2l4jhm} The same as Fig.~\ref{2l4jlm} but for $\sqrt{s}=3$~TeV.}
\end{figure}

Finally, let us consider a full simulation for the signatures $e^+e^- \to ZH \to ZZZ \to 6\ell$ and $e^+e^- \to ZH \to ZW^+W^- \to 2\ell + 4j$. The total cross section for the $6\ell$ signature is very low due to the small $Z \to 2\ell$ branching ratio. Scaling the Higgs production cross section from above we find that $\sigma(e^+e^-\to6\ell)\sim 0.1$~fb at most, so a very high luminosity is required for studying this channel. 
Our conclusion differs from that of \cite{Barger:2009xg}, where the $6\ell$ final state in $pp\to HZ$ was suggested as a promising channel to study $Z'$ resonances at the LHC. This happens as in our model the heavy vector states couple to the SM fermions only through mixing with the electroweak gauge bosons, and therefore their fermionic couplings are suppressed.
Hence we concentrate on the $2\ell + 4j$ signature, which has considerably larger cross section.
We adapt as acceptance cuts for charged leptons and jets:
\begin{eqnarray}\label{acceptancecuts}
|p_\perp^\ell| > 10~{\rm GeV},\qquad|p_\perp^j| &>& 20~{\rm GeV},\qquad |\cos\theta^{\ell,j}| < 0.95 \ .
\end{eqnarray}
Here $\theta^{\ell,j}$ is, as before, either the angle between the beam axis and the lepton or the beam axis and the jet.  We also require that all jet pairs are well separated:
\begin{equation}
 \Delta \eta^2 + \Delta \phi^2 > 0.3^2 \ ,
\end{equation}
where $\Delta \eta$ is the pseudorapidity difference between the total jet momenta and $\Delta \phi$ is the azimuthal separation. Also, the energies of the leptons and jets are smeared as explained above.

Event distributions for the signal  $e^+e^- \to HZ \to  2\ell + 4j$ at $\sqrt{s}=1$~TeV is presented in Fig.~\ref{2l4jlm} for the integrated luminosity of $100$~fb$^{-1}$ \footnote{We only included the dominant $H \to WW \to 4j$ decay. The signal cross section will be slightly enhanced if other channels are added.}. The dotted lines are our estimate for the background identified with the SM process $e^+e^- \to ZW^+W^- \to  2\ell + 4j$.
For comparison we also show the SM cross section for the same signature (dashed lines), i.e., $e^+e^- \to HZ \to  2\ell + 4j$ in the absence of heavy composite vectors in the intermediate state. 
We also repeated the study for $\sqrt{s}=3$~TeV and for higher mass of the composite states in Fig.~\ref{2l4jhm}.

\begin{table}
\begin{eqnarray}
\begin{array}{c|c|c||c|c}
 M_A(\textrm{GeV}) & \quad M_H(\textrm{GeV})    & \quad \ \tilde g \quad \mbox{}  &\quad L(\textrm{fb}^{-1})\ \textrm{for}\  3\sigma   &\quad L(\textrm{fb}^{-1})\  \textrm{for}\  5\sigma  \\
\hline\hline
750 &  \quad 200 &  \quad \ 2  \quad \mbox{} & \quad  18 &\quad 59  \\
750 &  \quad 200 & \quad \ 5 \quad \mbox{} & \quad  4.0 &\quad 12 \\
750 & \quad  400 &  \quad \ 2  \quad \mbox{} & \quad  80 &\quad 230  \\
750 & \quad  400 & \quad \ 5 \quad \mbox{} & \quad  7.8 &\quad 24 \\
\hline
950 & \quad  200 &  \quad \ 2  \quad \mbox{} & \quad 0.28 &\quad 0.86 \\
950 & \quad  200 & \quad \ 5 \quad \mbox{} & \quad  0.48 &\quad 1.5  \\ 
950 & \quad  400 &  \quad \ 2  \quad \mbox{} & \quad  0.61 &\quad 1.9 \\
950 & \quad  400 & \quad \ 5 \quad \mbox{} & \quad  0.71 &\quad 2.2  
\end{array} \nonumber
\end{eqnarray}
\caption{Estimates for required luminosities for $3\sigma$ and $5\sigma$ signals of the axial vector resonance in associated Higgs production at $\sqrt{s}=1$~TeV.}
\label{table:ss1est}
\end{table}

\begin{table}
\begin{eqnarray}
\begin{array}{c|c|c||c|c}
 M_A(\textrm{GeV}) & \quad M_H(\textrm{GeV})    & \quad \ \tilde g \quad \mbox{}  &\quad L(\textrm{fb}^{-1})\ \textrm{for}\  3\sigma   &\quad L(\textrm{fb}^{-1})\  \textrm{for}\  5\sigma  \\
\hline\hline
2250 & \quad  400 &  \quad \ 2  \quad \mbox{} & \quad  5.0 &\quad 12  \\
2250 & \quad  400 & \quad \ 5 \quad \mbox{} & \quad  56 &\quad 170 \\
\hline
2900 & \quad  400 &  \quad \ 2  \quad \mbox{} & \quad  5.2 &\quad 13 \\
2900 & \quad  400 & \quad \ 5 \quad \mbox{} & \quad  77 &\quad 210  
\end{array} \nonumber
\end{eqnarray}
\caption{Estimates for required luminosities for $3\sigma$ and $5\sigma$ signals of the axial vector resonance in associated Higgs production at $\sqrt{s}=3$~TeV.}
\label{table:ss3est}
\end{table}

Let us then estimate the luminosity which is necessary to study this process.  We use the following cuts 
\begin{eqnarray}\label{lastcuts}
 |M_{2\ell4j}-M_{R}| & < & \max\left(0.5\Gamma_{R},0.5 \sqrt{\frac{M_{R}}{\rm GeV}} {\rm GeV}\right) \ ,\nonumber \\
 |M_{4j}-M_{H}| & < & \max\left(0.5\Gamma_{H},0.5 \sqrt{\frac{M_{H}}{\rm GeV}} {\rm GeV}\right)
\end{eqnarray}
so that the four jets are produced by the Higgs decay. The significance of the signal is then calculated as detailed above. The luminosities corresponding to the distributions of Figs.~\ref{2l4jlm} and~\ref{2l4jhm} are listed in Tables~\ref{table:ss1est} and~\ref{table:ss3est}. We only considered the most significant axial vector peak which is $R_1$ in the $\sqrt{s}=1$~TeV scenario and $R_2$ for  $\sqrt{s}=3$~TeV.

\begin{figure}[thb!]
\includegraphics[width=0.45\textwidth]{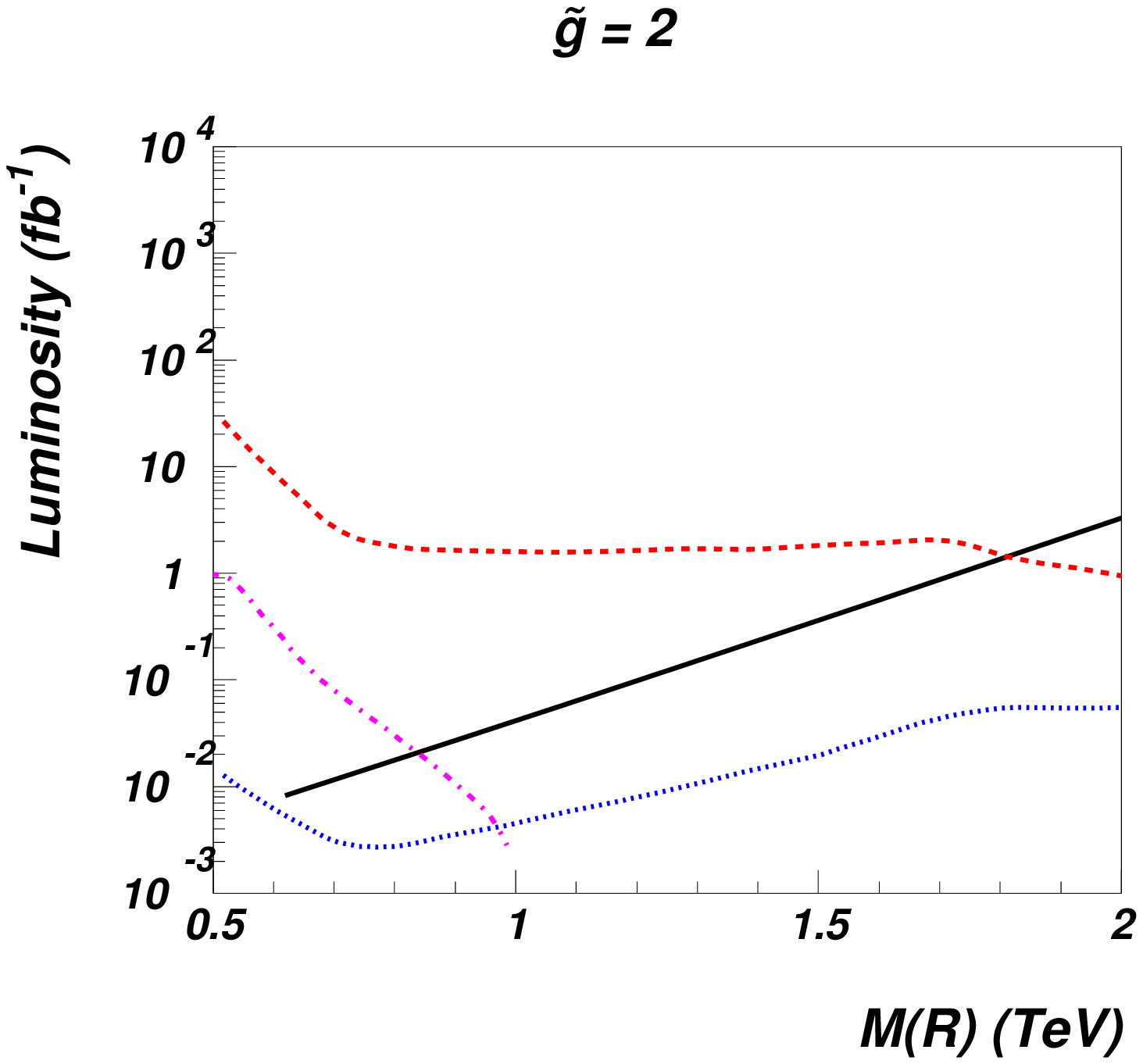}%
\includegraphics[width=0.45\textwidth]{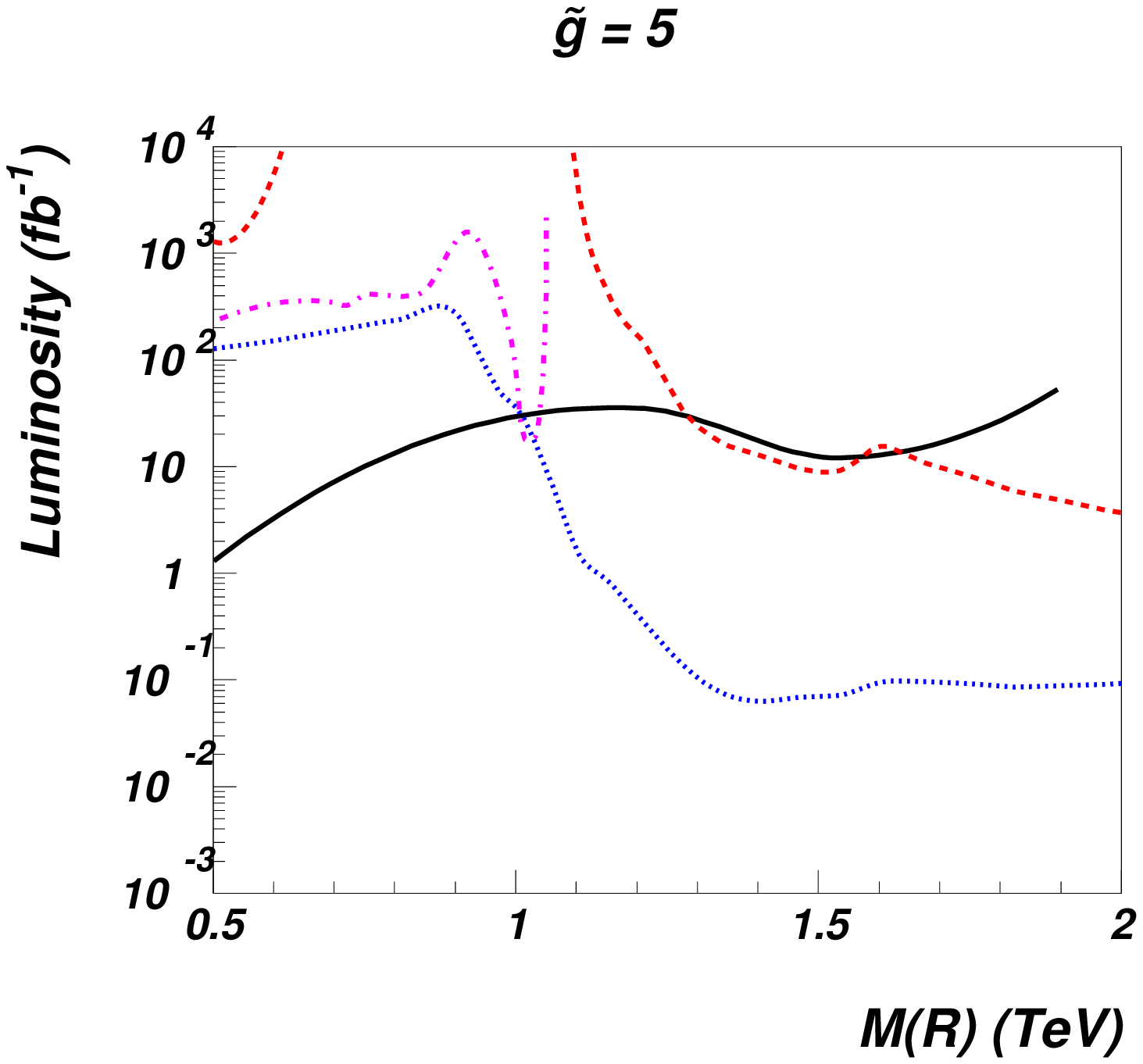}
\caption{\label{reach}Estimated luminosity required for a $5\sigma$ discovery of {\em any} vector resonance in {\em any} of the channels of Figs.~\ref{lumiss1}, \ref{lumiss3}, and~\ref{lumissvar}  at LCs as a function of the resonance mass. The solid black lines are the estimates for LHC \cite{Belyaev:2008yj},  dash-dotted magenta (dashed red) lines are for a 1~TeV (3~TeV) LC, and the dotted blue lines are for LC in the scanning mode ($\sqrt{s} = M(R)+30$~GeV). Left: $\tilde g=2$, right: $\tilde g=5$.}
\end{figure}

\section{Comparison to LHC and conclusions}

We combine the reach estimates of the signatures (1) and (2) (Figs.~\ref{lumiss1}, \ref{lumiss3}, and~\ref{lumissvar}) in Fig.~\ref{reach}, and compare them to the LHC reach \cite{Belyaev:2008yj}. We only present the luminosity required for a $5 \sigma$ discovery of the most significant resonance in either of the signatures ($e^+e^- \to R_{1,2}^0 \to \ell^+\ell^-$ and  $e^+e^- \to R_{1,2}^0 \to \nu+\ell+2j$)
for the LC curves. For the LHC at $\sqrt{s} = 14$~TeV the included channels are $pp \to R_{1,2}^0 \to \ell^+\ell^-$, $pp \to R_{1,2}^\pm \to \ell^\pm\nu$, and $pp \to R_{1,2}^\pm \to 3\ell+\nu$. Since the luminosity for LC  in the scanning mode (the dotted blue curve) is mostly below the LHC one (solid black), the LC can confirm, or actually discover, resonances found or missed at the LHC already at relatively low luminosities. We also observe that the LC working at fixed energy  improves the LHC reach when the resonance masses are near the beam energy. In particular, a 3~TeV LC can discover resonances at large values of their masses $\gtrsim 2$~TeV better than the LHC. We recall that heavy spin-one resonances, in the range of energy accessible to LCs, are able to delay the  unitarity violation of the WW scattering till several TeVs \cite{Foadi:2008ci}.
On the other hand these plots suggest that in the low mass region (below 1~TeV) for $\tilde g=5$ there is no improvement with respect to the LHC. However, we have not included the $e^+e^- \to R_{1,2} \to HZ$ channel for which the final state is more involved which could also improve the LC reach in this region of parameter space. In fact, the lightest (axial) vector resonance decays dominantly to $HZ$, and as seen from Table~\ref{table:ss1est}, including the signature (3) is expected to decrease the required LC luminosities by a factor of about $10^{2}$ for $\tilde{g} = 5$ and $M_A$ less than a TeV.

We have analyzed the reach of LCs for models of dynamical electroweak symmetry breaking and shown that they can be used to efficiently discover composite vector states with a fairly low luminosity.

\acknowledgments

We thank for discussions A.~Belyaev. MTF acknowledges a VKR Foundation Fellowship. MJ is supported in part by Regional Potential program of the E.U.FP7-REGPOT-2008-1: CreteHEPCosmo-228644 and by Marie Curie contract PIRG06-GA-2009-256487.

\end{document}